\documentclass[aps,preprint,showpacs,showkeys,nofootinbib,tightenlines]{revtex4}

\usepackage{graphicx}  %
\usepackage{amsmath}
\usepackage{bm}  %
\usepackage{ulem}
\usepackage[titletoc,title]{appendix}
\usepackage{comment}

\newcommand{\bea}{\begin{eqnarray}}
\newcommand{\eea}{\end{eqnarray}}
\newcommand{\beq}{\begin{equation}}
\newcommand{\eeq}{\end{equation}}
\newcommand{\bqa}{\begin{eqnarray}}
\newcommand{\eqa}{\end{eqnarray}}

\def\mqo2{{\!\!\!}}

\begin{document}

\title{
Harmonic and Subharmonic Association and Dissociation \\
of Universal Dimers in a Thermal Gas}
 
\author{Abhishek Mohapatra}
\email{mohapatra.16@buckeyemail.osu.edu}
\affiliation{Department of Physics,
         The Ohio State University, Columbus, OH\ 43210, USA\\}

\author{Eric Braaten}
\email{braaten@mps.ohio-state.edu}
\affiliation{Department of Physics,
         The Ohio State University, Columbus, OH\ 43210, USA\\}

\date{\today}
%\date{November 2007}

\begin{abstract}
In a gas of ultracold atoms 
whose scattering length is controlled by a magnetic Feshbach resonance, 
atoms can be associated into universal dimers 
by an oscillating magnetic field.
In addition to the harmonic resonance  with frequency near that 
determined by the dimer binding energy,
there is a subharmonic resonance with half that frequency.
If the thermal gas contains dimers,
they can be dissociated into unbound atoms 
by the oscillating magnetic field.
We show that the transition rates for association and dissociation
can be calculated by treating the oscillating magnetic field
as a sinusoidal time-dependent perturbation proportional to the contact operator.
Many-body effects are taken into account
through transition matrix elements of the contact operator.
We calculate both the harmonic and subharmonic transition rates analytically
for association in a thermal gas of atoms
and dissociation  in a thermal gas of dimers.
\end{abstract}

\smallskip
\pacs{31.15.-p, 34.50.-s, 67.85.Lm, 03.75.Nt, 03.75.Ss}
\keywords{
Fermi gases, scattering of atoms and molecules. }
\maketitle

%%%%%%%%%%%%%%%%%%%%%%%%%%%%%%%%%%%%%%%%
%                                      
%           Introduction               
%                                      
%%%%%%%%%%%%%%%%%%%%%%%%%%%%%%%%%%%%%%%%
              
\section{Introduction}

The use of {\it magnetic Feshbach resonances} to control the interaction 
strengths of ultracold atoms has led to significant advances in our
understanding of strong interactions in few-body and many-body physics.
The effects of time-dependent strong interactions can be studied by
using a time-dependent magnetic field.
A particularly interesting case is a sinusoidally modulated magnetic field.
Atoms can be associated into universal molecules composed of atoms with 
a large scattering length by modulating the magnetic field 
with a frequency near that determined by the binding energy of the molecule.
The measurement of the binding energy of a molecule
by the resonance in the oscillation frequency is called 
{\it magnetic-field modulation spectroscopy} 
or sometimes {\it wiggle spectroscopy}.

Modulation of the magnetic field was pioneered by Thompson, Hodby, and Wieman
to associate $^{85}$Rb atoms into dimers \cite{Wieman0505}.
Papp and Wieman used magnetic-field modulation spectroscopy to measure
the small binding energies of dimers composed 
of $^{85}$Rb  and $^{87}$Rb atoms \cite{Wieman0607}.
Weber et al.\ used a resonantly modulated magnetic field to associate  
$^{41}$K and $^{87}$Rb atoms into dimers and  
to measure their binding energies \cite{Inguscio0808}. 
They also observed subharmonic resonances at half the frequency 
determined by the binding energies of the dimers.
Lange et al.\ used  magnetic-field modulation spectroscopy to measure
the binding energies of $^{133}$Cs dimers \cite{Grimm0810}. 
Pollack et al.\ used a modulated magnetic field to excite
collective modes in a Bose-Einstein condensate of $^7$Li atoms \cite{Hulet1004}.
Machtey et al.\  used a modulated magnetic field
to associate $^7$Li atoms into Efimov trimers \cite{Khaykovich1201}.
Dyke, Pollack, and Hulet used magnetic-field modulation spectroscopy to measure
the binding energies of $^7$Li dimers in both a Bose-Einstein condensate 
and a thermal gas \cite{Hulet1302}.  
In the thermal gas, they also observed a subharmonic resonance.
Smith recently pointed out that a sinusoidally oscillating magnetic field
near a Feshbach resonance can also be used to control the scattering length,
and he showed that the resonance  parameters are universal functions 
of the magnetic field \cite{DHSmith1503}.

A theoretical treatment of the association of atoms into dimers 
by an oscillating magnetic field
was first presented by Hanna, K\"ohler, and Burnett in 2007 \cite{HKB0609}.
An alternative approach was recently developed by Brouard and Plata \cite{BP1503}.
Both groups described the two-atom system by a two-channel 
model consisting of a continuum of atom-pair states and a discrete
molecular state. 
They calculated the probability for the association of atom pairs into dimers 
as a function of time by solving the time-dependent Schr\"odinger equation
for the two coupled channels.   
In Ref.~\cite{BP1503},
some qualitative aspects of the harmonic and subharmonic
association processes were derived analytically.
The results for association probabilities from both groups were completely numerical.

A much simpler approach to this problem was recently introduced in Ref.~\cite{LSB:1406}.
It was inspired by Tan's {\it adiabatic relation}, 
which expresses the change in the energy $E$ of a system 
due to a change in the scattering length $a$
in terms of an extensive thermodynamic variable that is conjugate to $1/a$ 
called the {\it contact} $C$ \cite{Tan0508}.  
In the case of fermions with mass $m$ and two spin states, the adiabatic relation is
%-----------------
\begin{equation}
\frac{d~~~}{d(1/a)} E =- 
\frac{\hbar^2}{4 \pi m }C .
\label{E-C}
\end{equation}
%----------------- 
(In the case of identical bosons, the right side should be multiplied by $1/2$.)
Ref.~\cite{LSB:1406} pointed out that the transition rates for the association 
of atoms into universal dimers
can be calculated by treating the oscillating magnetic field
as a sinusoidal time-dependent perturbation proportional to the contact operator.
The association rates were calculated for a thermal gas of atoms 
and for a dilute Bose-Einstein condensate of atoms.
In this approach, many-body effects are taken into account
through transition matrix elements of the contact operator.

In this paper, we extend the approach of Ref.~\cite{LSB:1406}
to the dissociation rates of universal dimers
and to subharmonic transitions.
In Sections~\ref{sec:Transition},
we derive general formulas for the harmonic and subharmonic transition rates
using time-dependent perturbation theory.
In Sections~\ref{sec:harmassoc} and  \ref{sec:harmdissoc},
we calculate the leading harmonic contributions
to the association rate in a thermal gas of atoms 
and the dissociation rate in a thermal gas of dimers.
They come from a first-order perturbation in the contact operator.
In Sections~\ref{sec:subharmassoc} and  \ref{sec:subharmdissoc},
we calculate the dominant subharmonic contributions
to the association rate in a thermal gas of atoms 
and the dissociation rate in a thermal gas of dimers.
They come from a second-order perturbation in the contact operator.
In Section~\ref{sec:7Li}, we apply our results for the
association rate to a thermal gas of $^7$Li atoms.
In Section~\ref{sec:discsum}, we summarize previous theoretical 
treatments of association into dimers using a modulated magnetic field,
and we compare them with our results for association.

%\newpage

%%%%%%%%%%%%%%%%%%%%%%%%%%%%%%%%%%%%%%%%
%                                      
%           Transition Rate           
%                                      
%%%%%%%%%%%%%%%%%%%%%%%%%%%%%%%%%%%%%%%%

\section{Transition Rates}
\label{sec:Transition}

In this section, we derive general formulas for transition rates at first order 
and second order in time-dependent perturbation theory.
We focus on the case of fermionic atoms with equal mass $m$ and two spin states.
(We also give the corresponding results for identical bosons.)

%%%%%%%%%%%%%%%%%%%%%%%%%%%%%%%%%%%%%%%%%%%%%%%
\subsection{Perturbing Hamiltonians}
%%%%%%%%%%%%%%%%%%%%%%%%%%%%%%%%%%%%%%%%%%%%%%%

Near a magnetic Feshbach resonance, 
the scattering length $a$ of the atoms is a function of the magnetic field:
%-----------------
\begin{equation}
a(B) = a_{\rm bg} [1 - \Delta/(B - B_0)],
\label{a-B}
\end{equation}
%----------------- 
where $a_{\rm bg}$ is the background scattering length
and $B_0$ and $B_0 + \Delta$ are the positions of the pole 
and the zero of the scattering length, respectively. 
We consider a time-dependent magnetic field that has a constant value $\bar B$ for $t<0$
and is modulated with a small amplitude $b$ around the average value $\bar B$ for $t>0$:
%-----------------
\begin{eqnarray}
B(t) &=& \bar B \hspace{3cm} t<0,
\nonumber
\\
&=& \bar B + b \sin(\omega t) \hspace{1cm} t>0.
\label{oscillating}
\end{eqnarray}
%----------------- 
If the oscillating magnetic field is inserted into Eq.~\eqref{a-B},
it implies a time-dependent scattering length $a(t)$.

Tan's adiabatic relation in Eq.~\eqref{E-C} implies that the leading perturbation 
in the Hamiltonian  for $t>0$ is proportional  to the contact operator:
%-----------------
\begin{equation}
H(t) - H(0)  = 
- \frac{\hbar^2}{4 \pi m} \left( \frac{1}{a(t)} - \frac {1}{\bar a} \right) C,
\label{Hpert}
\end{equation}
%----------------- 
where $\bar a = a(\bar B)$ is the scattering length in the absence of the modulated 
magnetic field. 
In Appendix~\ref{appendix-1}, quantum field theory methods are used to argue 
that this is the only perturbation that contributes in the zero-range limit.
The inverse scattering length can be expanded in powers of $b$:
%-----------------
\begin{equation}
\frac{1}{a(t)} = 
\frac{1}{\bar a} 
- \frac{1}{a_{\rm bg}} \left(\frac{b\Delta}{(\Delta + B_0 - \bar B)^2} \right) \sin(\omega t) 
- \frac{1}{a_{\rm bg}} 
\left( \frac{b^2\Delta }{(\Delta + B_0 - \bar B)^3} \right) \sin^2(\omega t)+ \ldots .
\label{a-inverse}
\end{equation}
%----------------- 
The coefficients of the powers of $b$ have well-behaved 
limits as $\bar B$ approaches the Feshbach resonance at $B_0$.
Inserting the expansion in Eq.~\eqref{a-inverse} into Eq.~\eqref{Hpert},
we can identify terms in the perturbing Hamiltonian 
that are first and second order in $b$:
%-----------------
\begin{subequations}
\begin{eqnarray}
H_{\rm 1}(t) &=& 
\frac{\hbar^2}{4 \pi m a_{\rm bg}}
\left( \frac{b\Delta}{(\Delta + B_0 - \bar B)^2} \right)
C  \sin(\omega t),\\ 
H_{\rm 2}(t) &=& 
\frac{\hbar^2}{4 \pi m a_{\rm bg}}
\left( \frac{b^2\Delta}{(\Delta + B_0 - \bar B)^3} \right)
C  \sin^2(\omega t).
\label{H-C}
\end{eqnarray}
\end{subequations}
%----------------- 
(In the case of identical bosons, the right sides should be multiplied by $1/2$.)
If $|b| \ll |\Delta|$, the effects of $H_1$ and $H_2$
can be taken into account as time-dependent perturbations.
The first-order perturbation in $H_{\rm 1}$ drives transitions to states whose energies are 
higher or lower by $\hbar \omega$. 
We refer to such transitions as  {\it harmonic} transitions.
The first-order perturbation in $H_{\rm 2}$ 
and the second-order perturbation in $H_{\rm 1}$ both drive transitions 
to states whose energies  differ by 0 or $\pm 2\hbar\omega$.  
We refer to  transitions to states whose energies are 
higher or lower by $2\hbar \omega$ as {\it subharmonic}  transitions.

%%%%%%%%%%%%%%%%%%%%%%%%%%%%%%%%%%%%%%%%%%%%%%%
\subsection{Fermi's Golden Rule}
%%%%%%%%%%%%%%%%%%%%%%%%%%%%%%%%%%%%%%%%%%%%%%%

We first consider transitions from the first-order perturbation in $H_{\rm 1}$. 
We take the initial state $|i\rangle$ to be an energy eigenstate with energy $E_{i}$.
We consider the transition to a distinct energy eigenstate $|f\rangle$ with energy $E_{f}$.
At first order in perturbation theory, 
the probability amplitude for the final state $|f\rangle$ at time $T$ is 
%----------------- 
\begin{equation}
 a_{f}^{(1)}(T)=
 \frac{i\hbar}{8\pi m a_{\rm bg}}
 \left( \frac{b\Delta }{(\Delta + B_0 - \bar B)^2} \right)
 \left[\frac{e^{i(\omega_{ fi}+\omega)T}-1}{\omega_{ fi}+\omega}-
 \frac{e^{i(\omega_{ fi}-\omega)T}-1}{\omega_{ fi}-\omega}\right]
 \langle f|C|i\rangle,
\label{coefficient-1}
\end{equation}
%----------------- 
where $\omega_{ fi}= (E_{ f}-E_{ i})/\hbar$.
The two terms inside the brackets 
have absolute values that increase linearly with $T$ 
in the limits $\omega_{fi} \to -\omega$ and $\omega_{fi} \to +\omega$, respectively.  
By applying Fermi's Golden Rule, we obtain the transition rate 
summed over final states $| f \rangle$:
%-----------------
\begin{equation}
\Gamma_{1}^{(1)}(\omega)  = 
\frac{\hbar^2 }{64 \pi^2 m^2 a_{\rm bg}^2 }
\left(\frac{b\Delta }{(\Delta + B_0 - \bar B)^2}\right)^2
\sum_f \big| \langle f | C | i \rangle \big|^2
\sum_\pm 2 \pi \delta(\omega_{fi}\pm\omega).
\label{Gamma1}
\end{equation}
%----------------- 
(In the case of identical bosons, the prefactor should be multiplied by $1/4$.) 
This transition rate is non-zero only for final states whose energy 
differs from $E_{i}$ by $\pm\hbar\omega$, so
it contributes to the harmonic transition rate $\Gamma_1(\omega)$.
The superscript (1) on $\Gamma_{1}^{(1)}$
indicates that it comes from the first-order perturbation in $H_1$.

We next consider transitions from the first-order perturbation in $H_{2}$. 
At first order in perturbation theory, 
the probability amplitude for the final state $|f\rangle$ at time $T$ is 
%-----------------
\begin{equation}
 a_{f}^{(2)}(T)=
 \frac{\hbar}{16\pi m a_{\rm bg}}
\left( \frac{b^2\Delta }{(\Delta + B_0 - \bar B)^3} \right)
 \left[\frac{e^{i(\omega_{ fi}+2\omega)T}-1}{\omega_{ fi}+2\omega}+
 \frac{e^{i(\omega_{ fi}-2\omega)T}-1}{\omega_{ fi}-2\omega}+\ldots\right]
 \langle f|C|i\rangle.
\label{coefficient-2}
\end{equation}
%-----------------
Inside the brackets, we have shown explicitly only those terms 
whose absolute values increase linearly 
with $T$ in the limits $\omega_{fi} \to \pm2\omega$. 
By applying Fermi's Golden Rule, we obtain the transition rate 
summed over final states $| f \rangle$:
%-----------------
\begin{equation}
\Gamma_{2}^{(2)}(\omega)  = 
\frac{\hbar^2 }{256 \pi^2 m^2 a_{\rm bg}^2 }
\left(\frac{b^2\Delta }{(\Delta + B_0 - \bar B)^3}\right)^2
\sum_f \big| \langle f | C | i \rangle \big|^2
\sum_\pm 2 \pi \delta(\omega_{fi}\pm2\omega).
\label{Gamma2-2}
\end{equation}
%----------------- 
(In the case of identical bosons, the prefactor should be multiplied by $1/4$.) 
This transition rate is nonzero only for final states whose energy 
differs from $E_{i}$ by $\pm2\hbar\omega$,
so it contributes to the subharmonic transition rate $\Gamma_2(\omega)$.
The superscript (2) on $\Gamma_{2}^{(2)}$
indicates that it comes from the first-order perturbation in $H_2$.
The subharmonic transition rate in Eq.~\eqref{Gamma2-2} is determined by
the same transition matrix element $\langle f | C | i \rangle$
of the contact operator as the harmonic transition rate in Eq.~\eqref{Gamma1}.
It can be expressed in terms of the leading harmonic transition rate
at twice the frequency:
%-----------------
\begin{equation}
\Gamma_{2}^{(2)}(\omega)  = 
\frac14
\left(\frac{b}{\Delta + B_0 - \bar B}\right)^2
\Gamma_1^{(1)}(2\omega).
\label{Gamma2-2:simple}
\end{equation}
%-----------------

Finally we consider transitions from the second-order perturbation in $H_{\rm 1}$.
At second order in perturbation theory, 
the probability amplitude for the final state $|f\rangle$ at time $T$ is
%-----------------
\begin{eqnarray}
a_{f}^{(1,1)}(T)&=&
-\frac{\hbar^2}{64\pi^2m^2 a_{\rm bg}^2}
\left( \frac{b \Delta}{(\Delta + B_0 - \bar B)^2} \right)^2
\sum_{m\ne i} \langle f|C|m\rangle \langle m|C|i\rangle
\nonumber\\
&&\hspace{2.5 cm}\times
\left[ \frac{e^{i(\omega_{fi}+2\omega)T}-1}{(\omega_{fi}+2\omega)(\omega_{mi}+\omega)}
+\frac{e^{i(\omega_{fi}-2\omega)T}-1}{(\omega_{fi}-2\omega)(\omega_{mi}-\omega)}
+\ldots\right],
\label{coefficient-11}
\end{eqnarray}
%-----------------
where the sum is over intermediate states $|m\rangle$ distinct from $|i\rangle$. 
Inside the brackets, we have shown explicitly only those terms 
whose absolute values increase linearly with $T$ if $\omega_{fi}$ is $\pm2\omega$. 
By applying Fermi's Golden Rule, we obtain the transition rate 
summed over final states $| f \rangle$:
%-----------------
\begin{equation}   
\Gamma_{2}^{(1,1)}(\omega)  = 
\frac{\hbar^4 }
    {4096 \pi^4 m^4 a_{\rm bg}^4 }\left(\frac{b\Delta }{(\Delta + B_0 - \bar B)^2}\right)^4
\sum_f \sum_\pm \left| \sum_{m\ne i} 
\frac{\langle f | C | m \rangle\langle m |C|i\rangle}{\omega_{mi}\pm\omega} \right|^2
2 \pi \delta(\omega_{fi}\pm2\omega).
\label{Gamma2-11}
\end{equation}
%-----------------
(In the case of identical bosons, the prefactor should be multiplied by 1/16.) 
This transition rate is nonzero only for final states whose energy 
differs from $E_{i}$ by $\pm2\hbar\omega$,
so it contributes to the subharmonic transition rate $\Gamma_2(\omega)$.
The superscript (1,1) on $\Gamma_{2}^{(1,1)}$
indicates that it comes from the second-order perturbation in $H_1$.
There is an additional factor of $1/a_{\rm bg}^2$ in the prefactor for $\Gamma_{2}^{(1,1)}$ 
compared to $\Gamma_{2}^{(2)}$.  The relative importance of these two
contributions is determined by the canceling length scales provided by the 
contact matrix elements and the frequency denominator.
If $\Gamma_{2}^{(1,1)}$ and $\Gamma_{2}^{(2)}$ have the same order of magnitude, 
the interference between the first-order perturbation in $H_2$ 
and the second-order perturbation in $H_1$ would have to be taken into account.
By explicit calculations of subharmonic transition rates in a thermal gas,
we will find that the additional dimensionless factor in $\Gamma_{2}^{(1,1)}$
is $(\bar a/a_{\rm bg})^2$.  Thus $\Gamma_{2}^{(1,1)}$ is much larger than
$\Gamma_{2}^{(2)}$ if $\bar B$ is near a Feshbach resonance.

%%%%%%%%%%%%%%%%%%%%%%%%%%%%%%%%%%%%%%%%%%%%%%%
\subsection{Thermal System}
%%%%%%%%%%%%%%%%%%%%%%%%%%%%%%%%%%%%%%%%%%%%%%%

The transitions rates in Eqs.~\eqref{Gamma1}, 
\eqref{Gamma2-2}, and \eqref{Gamma2-11} apply to an initial state $|i \rangle$
that is an energy eigenstate.  
A thermal system is described instead by a density matrix.
For a completely thermalized system, the density matrix is
$\rho=  \exp(- \beta H)/{\rm Tr}( \exp(- \beta H))$,
where $H$ is the Hamiltonian and  $\beta = 1/k_BT$.
By expressing the modulus-squared of an amplitude
as the product of the amplitude and its complex conjugate,
the dependence on the initial state in Eqs.~\eqref{Gamma1},
\eqref{Gamma2-2}, and \eqref{Gamma2-11} can be put in the form of 
the projection operator $| i \rangle \langle i |$
multiplied by a function $F(E_i)$ of the initial energy 
that includes the frequency delta function.
If the density matrix $\rho$ is diagonal in an energy basis,
the transition rate is obtained by making the substitution
%-----------------
\begin{equation}   
F(E_i) \, |i \rangle \langle i| \longrightarrow 
\sum_i F(E_i) \, |i \rangle \langle i | \rho | i \rangle \langle i|.
\label{|i><i|-rho}
\end{equation}
%-----------------

%%%%%%%%%%%%%%%%%%%%%%%%%%%%%%%%%%%%%%%%%%%%%%%
\subsection{Homogeneous System}
%%%%%%%%%%%%%%%%%%%%%%%%%%%%%%%%%%%%%%%%%%%%%%%

The contact operator $C$ is an extensive variable. It can be expressed
as the integral over space of the {\it contact density} operator:
%-----------------
\begin{equation}
C = \int d^3r \, {\cal C}(\bm{r}).
\label{C-density} 
\end{equation}
%----------------- 
If the initial and final states are homogeneous systems,
we can simplify the transition rates by expressing them in terms of 
matrix elements of the contact density operator.

The harmonic  transition rate $\Gamma_1^{(1)}(\omega)$ in Eq.~\eqref{Gamma1}
and the subharmonic transition rate  $\Gamma_2^{(2)}(\omega)$ in Eq.~\eqref{Gamma2-2} 
involve the factor $| \langle f | C | i \rangle |^2$.
By inserting the expression for $C$ in Eq.~\eqref{C-density}, 
we obtain matrix elements of the contact density at two different positions.
We can use translational invariance to put both operators at the same position $\bm r$.
One of the integrals over space then gives a momentum-conserving delta function.
The resulting expression for the modulus-squared of the transition matrix element is
%-----------------
\begin{equation}
\sum_f \big| \langle f | C | i \rangle \big|^2 =
\sum_f (2 \pi)^3 \delta^3(\bm{K}_f - \bm{K}_i)
\int \! d^3r  \,
\big| \langle f |{\cal C}(\bm r) | i \rangle \big|^2,
\label{homo-AA}
\end{equation}
%----------------- 
where  $\bm{K}_{i}$ and $\bm{K}_{f}$ are the total wave vectors of the initial 
and final states of the homogeneous system, respectively.
Homogeneity implies that $\big| \langle f |{\cal C}(\bm r) | i \rangle \big|^2$
is independent of the position $\bm r$.  Thus the integral $\int \! d^3r$
in Eq.~\eqref{homo-AA} just gives a factor of the volume $V$.

The subharmonic transition rate  $\Gamma_2^{(1,1)}(\omega)$ in Eq.~\eqref{Gamma2-11}
involves the product of four matrix elements of $C$.
By inserting the expression for $C$ in Eq.~\eqref{C-density}, 
we obtain matrix elements of the contact density at four different positions.
We can use translational invariance to put all four operators at the same position $\bm r$.
Three of the integrals over space then give momentum-conserving delta functions.
The resulting expression for the  factor in Eq.~\eqref{Gamma2-11}
that involves matrix elements of $C$ is
%----------------- 
\begin{eqnarray}
\sum_f \left| \sum_{m\ne i}
\frac{\langle f | C | m \rangle\langle m |C|i\rangle}{\omega_{mi}\pm\omega} \right|^2
&=& \sum_f (2\pi)^3\delta^3(\bm{K}_f-\bm{K}_i) \int d^{3}r
 \nonumber\\
&& 
\times
\sum_{m\ne i} (2\pi)^3  \delta^{3}(\bm{K}_{m}-\bm{K}_{i})\
\frac{\langle f|{\cal C}(\bm r) | m \rangle \langle m |{\cal C}(\bm r)|i\rangle}
      {\omega_{mi} \pm \omega}\,
 \nonumber\\
&&  
\times \sum_{m' \ne i}  (2\pi)^3  \delta^3(\bm{K}_{m' }-\bm{K}_i)
\frac{\langle i|{\cal C}(\bm r)|m' \rangle\langle m' |{\cal C}(\bm r)|f\rangle}
       {\omega_{m' i} \pm \omega},
\label{homo-D}
\end{eqnarray}
%----------------- 
where $\bm{K}_{m}$ and $\bm{K}_{m' }$ are the total momenta 
of the intermediate states $|m\rangle$ and $|m' \rangle$, respectively.
Homogeneity implies that the integrand is independent of the position $\bm r$.  
Thus the integral $\int \! d^3r$ just gives a factor of the volume $V$.

%%%%%%%%%%%%%%%%%%%%%%%%%%%%%%%%%%%%%%%%%%%%%%%
\subsection{Local Density Approximation}
%%%%%%%%%%%%%%%%%%%%%%%%%%%%%%%%%%%%%%%%%%%%%%%

For a many-body system whose number density varies slowly with the position $\bm{r}$,
the transition rate can be  simplified by using the {\it local density approximation}.
The transition rate is an extensive quantity.
For a homogeneous system, the expressions obtained by inserting
Eq.~\eqref{homo-AA} or Eq.~\eqref{homo-D} into the transition rate 
have an explicit factor of the volume $\int d^3r=V$.
If the transition rate is also proportional to the total number $N_i$ of some type of particle,
the additional factor must be the intensive combination $N_i/V$.
For a homogeneous system consisting of fermionic atoms with spin states 
1 and 2, the association rate is proportional to $N_1N_2$.
The local density approximation for the association rate 
in a system with local number densities $ n_1(\bm{r})$ and $n_2(\bm{r})$ 
can be obtained by making the substitution
%-----------------
\begin{equation}
N_1 N_2/V \longrightarrow
\int d^3r \, n_1(\bm{r}) n_2(\bm{r}).
\label{LDA-AA}
\end{equation}
%----------------- 
For a homogeneous system consisting of dimers, 
the disssociation rate is proportional to their total number $N_{\rm D}$.
The local density approximation for the disssociation rate 
in a system with local number density $n_{\rm D}(\bm{r})$ 
can be obtained by making the substitution
%-----------------
\begin{equation}
N_{\rm D}  \longrightarrow
\int d^3r \, n_{\rm D}(\bm{r}).
\label{LDA-D}
\end{equation}
%----------------- 

%\newpage

%%%%%%%%%%%%%%%%%%%%%%%%%%%%%%%%%%%%%%%%
%                                      
%           Harmonic Association Rate   
%                                      
%%%%%%%%%%%%%%%%%%%%%%%%%%%%%%%%%%%%%%%%

\section{Harmonic Association Rate}
\label{sec:harmassoc}

A pair of atoms with a large positive scattering length  can be associated 
into a universal dimer by an oscillating magnetic field.
In this section, we calculate the harmonic association rate 
in a thermal gas of atoms.
We also give the subharmonic association rate from first-order
perturbation theory.
We consider a gas of atoms that is in thermal equilibrium at temperature $T$.
For simplicity, we take the number densities
$n_1$ and $n_2$ of the atoms to be sufficiently low that their distributions are
given by Boltzmann statistics instead of Fermi-Dirac statistics.

%%%%%%%%%%%%%%%%%%%%%%%%%%%%%%%%%%%%%%%%%%%%%%%
\subsection{Initial and Final States}
%%%%%%%%%%%%%%%%%%%%%%%%%%%%%%%%%%%%%%%%%%%%%%%

We first consider a homogeneous gas consisting of $ N_1$ atoms of spin state 1 
and $N_2$ atoms of spin state 2 in a volume $V$. 
The two spin states interact with a large positive scattering length $\bar a$.
The universal dimer has a small binding energy $\hbar^2/m \bar a^2$.
For a gas in thermal equilibrium, 
the harmonic transition rate is given by Eq.~\eqref{Gamma1}
with the substitution in Eq.~\eqref{|i><i|-rho},
where $\rho=\rho_{\rm gas}$ is the density matrix for the thermal gas of atoms.
To simplify the presentation,
we will temporarily ignore the frequency delta function,
which depends on the energy $E_i$ of the states in the density matrix.
The terms in  Eq.~\eqref{Gamma1} that depend on the 
contact  operator can then  be expressed 
compactly as $\sum_{f}\langle f|C\rho_{\rm gas}C|f\rangle$.
We will insert the  frequency delta function at the end of the calculation.

In the low-density limit where 3-body and higher-body correlations  can be neglected, 
the density matrix $\rho_{\rm gas}$ 
can be expressed in terms of the density matrix $\rho_{\rm pair}$ 
for a pair of atoms in thermal equilibrium:
%-----------------
\begin{equation}
\sum_{f}\langle f|C\rho_{\rm gas}C|f\rangle = 
N_1N_2\sum_ {f}\langle f|C\rho_{\rm pair}C|f\rangle.
\label{density-matrix-gas}
\end{equation} 
%-----------------
The factor $N_1 N_2$ is the number of pairs of fermions in the two spin states.
(For a gas of $N$ identical bosons, the number of pairs  is $N^2/2$.)
The pair density matrix is normalized:  Tr$(\rho_{\rm pair}) = 1$.
On the left side of Eq.~\eqref{density-matrix-gas}, 
the sum over $f$ is over many-body final states 
that include a single dimer.  On the right side,
the sum over $f$ is over two-atom final states that consist of a single dimer.
The density matrix for a pair of atoms in thermal equilibrium is
%-----------------
\begin{equation}
\rho_{\rm pair} =  \frac{\lambda_{\rm T}^6}{V^2} 
\int_{\bm{K}}
\int_{\bm{k}}
\exp(-\beta\hbar^2K^2/4m -\beta\hbar^2k^2/m)
|\bm{K}, \bm{k}\rangle\langle \bm{k}, \bm{K}|,
\label{rho-pair}
\end{equation}
%-----------------
where $\beta = 1/k_BT$ and
$\lambda_{\rm T}$ is the thermal deBroglie wavelength for an atom with mass $m$:
%-----------------
\begin{equation}
\lambda_{\rm T} = \sqrt{2 \pi \hbar^2/m k_{\rm B} T}.
\label{lambda_T}
\end{equation}
%-----------------
The two-atom states $|\bm{K}, \bm{k}\rangle$ 
in Eq.~\eqref{rho-pair} are labeled by the 
center-of-mass wave vector $\bm{K} = \bm{k}_1+\bm{k}_2$
and the relative wave vector $\bm{k}=(\bm{k}_1-\bm{k}_2)/2$.
The integrals over the wave vectors are defined by
%-----------------
\begin{equation}
 \int_{\bm{k}}\equiv\int\frac{ d^3k}{(2\pi)^3}.
\label{momenum-integral}
\end{equation}
%-----------------
The wave vector states have delta-function normalizations: 
$\langle\bm{k}' |\bm{k}\rangle=(2\pi)^3\delta^3(\bm{k}'-\bm{k})$.
In the case $\bm{k}' = \bm{k}$, the infinite norm can be expressed 
as a factor of the volume:  $\langle\bm{k} |\bm{k}\rangle=V$.
The energy of a pair of atoms in the state $|\bm{K}, \bm{k}\rangle$  is
%-----------------
\begin{equation}
 E_{\rm AA}= \hbar^2K^2/4m + \hbar^2k^2/m.
 \label{E-AA}
\end{equation}
%-----------------

The sum over final states on the right hand side of Eq.~\eqref{density-matrix-gas} 
can be expressed as an integral over the wave vector ${\bm k}_{\rm D}$ of a dimer: 
%-----------------
\begin{equation}
\sum_f\langle f|C\rho_{\rm pair}C|f\rangle=
\int_{\bm{k}_{\rm {D}}}\langle {\bm k}_{\rm D}|C\rho_{\rm pair}C|{\bm k}_{\rm D}\rangle.
\label{comp-final-state-therm}
\end{equation}
%-----------------
The energy of the dimer is 
%-----------------
\begin{equation}
E_{\rm D} = -\hbar^2/m \bar{a}^2 + \hbar^2 k_{\rm D}^2/4m.
\label{E-D}
\end{equation}
%-----------------

%%%%%%%%%%%%%%%%%%%%%%%%%%%%%%%%%%%%%%%%%%%%%%%
\subsection{Matrix Elements}
%%%%%%%%%%%%%%%%%%%%%%%%%%%%%%%%%%%%%%%%%%%%%%%

Because the system is homogeneous, 
the analog of Eq.~\eqref{homo-AA} can be used to express the  
contact operators $C$ on the right side of Eq.~\eqref{comp-final-state-therm}
in terms of contact density operators at the same position $\bm{r}$.
The wave vector delta function in Eq.~\eqref{homo-AA} 
reduces to $\delta^3(\bm{k}_{\rm D} - \bm{K})$, 
and it can be used to integrate over $\bm{k}_{\rm D}$. 
The frequency delta function in Eq.~\eqref{Gamma1} reduces to  
%-----------------
\begin{equation}
 \sum_{\pm}2\pi\delta\big( (E_{\rm D} - E_{\rm AA})/\hbar \pm \omega\big)= 
 2\pi\delta \big(\omega - \hbar/m \bar{a}^2  - \hbar k^2/m \big).
\label{delta-omega:plus}
\end{equation}
%-----------------
In the sum over $\pm\omega$, only the $+\omega$  term contributes. 

The expression for the transition rate has been reduced to matrix elements
of the contact density operator of the form 
$\langle {\bm k}_{\rm D}|{\cal C}({\bm r})|{\bm K}, {\bm k}\rangle$. 
The matrix element is calculated in Appendix~\ref{appendix-2}, 
and is given by Eq.~\eqref{matrix-2}:
%----------------
\begin{eqnarray}
\langle {\bm k}_{\rm D}| {\cal C}(\bm r) |{\bm{K}}, {\bm{k}}\rangle
=\frac{\sqrt{128\pi^3 \bar a}}{1- i \bar a k}.
\label{<D|C|AA>}
\end{eqnarray}
%-----------------
The Gaussian integral over $\bm{K}$ can be evaluated analytically. 
The sum over final states of the matrix element in Eq.~\eqref{comp-final-state-therm}
 reduces to
%-----------------
\begin{equation}
\sum_{f} \langle f | C\rho_{\rm gas} {C^{\dagger}} | f \rangle =
128 \sqrt{2} \pi \bar{a} \lambda_{\rm T}^3\frac{N_1N_2}{V}
\int_{0}^{\infty}\hspace{-0.2cm}dk\frac{k^2}{1+k^2\bar{a}^2}
\exp(-\beta\hbar^2k^2/m).
\label{contact-trans-conti}
\end{equation}
%-----------------
Before integrating over $k$,
this must be multiplied by the frequency delta function in 
Eq.~\eqref{delta-omega:plus}.

%%%%%%%%%%%%%%%%%%%%%%%%%%%%%%%%%%%%%%%%%%%%%%%
\subsection{Harmonic Association Rate}
%%%%%%%%%%%%%%%%%%%%%%%%%%%%%%%%%%%%%%%%%%%%%%%

Our final result for the harmonic association rate $\Gamma_{1}^{(1)}(\omega)$ 
in the homogeneous gas can be obtained 
from Eq.~\eqref{Gamma1} by replacing $\sum_f |\langle f|C|i\rangle|^2$
by the right side of Eq.~\eqref{contact-trans-conti},
replacing the sum of frequency delta functions
by the right side of Eq.~\eqref{delta-omega:plus}, 
and  then using the delta function to integrate over $k$.
The local density approximation can be implemented by making the substitution 
for $N_1 N_2/V$ in  Eq.~\eqref{LDA-AA}. 
The threshold angular frequency for association is $\hbar/m \bar a^2$:
the emission of a smaller energy
from a pair of atoms is not enough to allow a transition to dimer.
For $\omega>\hbar/m \bar a^2$, the harmonic association  rate is 
%-----------------
\begin{equation}
\Gamma_{1}^{(1)}(\omega)  = 
\frac{2 \sqrt{2} \hbar^2 }{m^2 a_{\rm bg}^2 \bar{a}}
\left(\frac{b\Delta }{(\Delta + B_0 - \bar B)^2}\right)^2
\left( \int \! d^3r \, n_{1} (\bm{r}) n_{2}(\bm{r}) \right)
\frac{\lambda_{\rm T}^3  \kappa(\omega)}{\omega} 
\exp(-\beta\hbar^2 \kappa^2(\omega)/m),
\label{Gamma-thermal}
\end{equation}
%----------------- 
where 
%----------------- 
\begin{equation}
\kappa(\omega)= \sqrt{m \omega/\hbar - 1/\bar a^2}.
\label{omega-k}
\end{equation}
%----------------- 
(The harmonic association rate in a thermal gas of identical bosons 
with large scattering length was calculated in Ref.~\cite{LSB:1406}. 
It can be obtained from Eq.~\eqref{Gamma-thermal}
by replacing $n_{\rm 1}(\bm{r})n_{\rm 2}(\bm{r})$ by $n^2(\bm{r})/2$, 
where $n(\bm{r})$ is the local number density of identical bosons.)
If $k_BT \ll \hbar^2/m \bar a^2$, the harmonic association  rate 
in Eq.~\eqref{Gamma-thermal} has a narrow peak
with a maximum when $\omega$ is above 
the threshold $\hbar/m \bar a^2$ by approximately $k_BT/2\hbar$.
For large frequency, the  rate decreases 
as $\exp(- \hbar \omega/k_BT)$.

%%%%%%%%%%%%%%%%%%%%%%%%%%%%%%%%%%%%%%%%%%%%%%%
\subsection{First-Order Subharmonic Association Rate}
%%%%%%%%%%%%%%%%%%%%%%%%%%%%%%%%%%%%%%%%%%%%%%%

According to Eq.~\eqref{Gamma2-2:simple}, 
the contribution $\Gamma_{2}^{(2)}(\omega)$ to the subharmonic association rate 
from the first-order perturbation in $H_{\rm 2}$
can be expressed in terms of the harmonic association rate 
in Eq.~\eqref{Gamma-thermal} at twice the frequency.
The threshold angular frequency for subharmonic association 
is $\frac12(\hbar/m \bar a^2)$.
For $\omega>\frac12 (\hbar/m \bar a^2)$, the subharmonic association rate is 
%-----------------
\begin{equation}
\Gamma_{2}^{(2)}(\omega)  = 
\frac{ \sqrt{2} \hbar^2 }{4 m^2 a_{\rm bg}^2 \bar{a}}
\left(\frac{b^2\Delta}{(\Delta + B_0 - \bar B)^3}\right)^2
\left( \int \! d^3r \, n_{1} (\bm{r}) n_{2}(\bm{r}) \right)\\
\frac{\lambda_{\rm T}^3  \kappa(2\omega)}{\omega} 
\exp(- \beta\hbar^2 \kappa^2(2\omega)/m), 
\label{sub-transition-asso}
\end{equation}
%-----------------
where $\kappa(2\omega)$ is the function defined in Eq.~\eqref{omega-k} 
with $\omega$ replaced by $2\omega$:
%-----------------
\begin{equation}
\kappa(2\omega)= \sqrt{2m\omega/\hbar - 1/\bar a^2}.
\label{two-omega-k}
\end{equation}
%-----------------
If $k_BT \ll \hbar^2/m \bar a^2$, this contribution  
to the subharmonic association rate has a narrow peak
with a maximum when $\omega$ is above 
the threshold $ \frac12 (\hbar/m \bar a^2)$ by approximately $k_BT/4\hbar$.
The height of the peak is smaller than that of the harmonic association rate
by the factor $[b/(\Delta + B_0 - \bar B)]^2/4$.
   
%\newpage

%%%%%%%%%%%%%%%%%%%%%%%%%%%%%%%%%%%%%%%%
%                                      
%           Harmonic Dissociation Rate
%                                      
%%%%%%%%%%%%%%%%%%%%%%%%%%%%%%%%%%%%%%%%

\section{Harmonic Dissociation Rate}
\label{sec:harmdissoc}

A universal dimer can be dissociated by an oscillating magnetic field into
its constituent atoms.
In this section, we calculate the harmonic dissociation rate in a thermal gas 
of dimers.
We also give the subharmonic dissociation rate from first-order
perturbation theory.
We consider a gas of dimers in thermal equilibrium at temperature $T$. 
For simplicity, we take the number density $n_{\rm D}$ of dimers 
to be sufficiently low that their distribution is given by Boltzmann statistics 
instead of Bose-Einstein statistics.

%%%%%%%%%%%%%%%%%%%%%%%%%%%%%%%%%%%%%%%%%%%%%%%
\subsection{Initial and Final States}
%%%%%%%%%%%%%%%%%%%%%%%%%%%%%%%%%%%%%%%%%%%%%%%

We first consider a homogeneous gas 
consisting of $N_{\rm D}$ dimers in a volume $V$. 
If the gas is in thermal equilibrium, 
the harmonic transition rate is given by Eq.~\eqref{Gamma1}
with the substitution in Eq.~\eqref{|i><i|-rho},
where $\rho_{\rm gas}$ is the density matrix for the thermal gas of dimers. 
To simplify the presentation,
we will temporarily ignore the frequency delta function,
which depends on the energy $E_i$ of the states in the density matrix.
The terms in  Eq.~\eqref{Gamma1} that depend on the 
contact  operator can be expressed 
compactly as $\sum_{f}\langle f|C\rho_{\rm gas}C|f\rangle$.
We will insert the  frequency delta function at the end of the calculation.

In the low-density limit where correlations between dimers can be neglected,
the density matrix $\rho_{\rm gas}$ can be expressed 
in terms of the density matrix $\rho_{\rm dimer}$ for a single dimer in thermal equilibrium:
%----------------- 
\begin {equation}
\sum_{f}\langle f|C\rho_{\rm gas}C|f\rangle = 
N_{\rm D}\sum_{f}\langle f|C\rho_{\rm{dimer}}C|f\rangle.
 \label{density-matrix-dimergas} 
\end {equation}
%----------------- 
The dimer density matrix is normalized:  Tr$(\rho_{\rm dimer}) = 1$.
On the left side of Eq.~\eqref{density-matrix-dimergas}, 
the sum over $f$ is over many-body final states 
that include an unbound pair of atoms.  On the right side, 
the sum over $f$ is over two-atom final states that consist of an unbound pair of atoms.
The density matrix for a dimer in thermal equilibrium is
%----------------- 
\begin{equation}
\rho_{\rm dimer}= \frac{\lambda_T^3}{2\sqrt{2}V}\int_{\bm{k}_{\rm D}}
\exp\left(-\beta\hbar^2k_{\rm D}^2/4m\right)\hspace{0.1cm}|{\bm k}_{\rm D}
\rangle\langle{\bm k}_{\rm D}|,
\label{rho-dimer}
\end{equation}
%----------------- 
where $\lambda_{\rm T}$ is the thermal deBroglie wavelength for an atom
in Eq.~\eqref{lambda_T}. 
The energy $E_{\rm D}$ of the dimer is given in Eq.~\eqref{E-D}.

The sum over final states on the right side of Eq.~\eqref{density-matrix-dimergas} 
can be expressed as integrals over the total wave vector and the
relative wave vector of a pair of atoms:
%----------------- 
\begin{equation}
\sum_f\langle f|C\rho_{\rm dimer}C|f\rangle=
\int_{\bm{K}}\int_{\bm{k}}\hspace{0.1 cm}
\langle {\bm K}, {\bm{k}}|C\rho_{\rm dimer}C|{\bm K}, {\bm{k}}\rangle.
\label{comp-final-state-atom} 
\end{equation}
%----------------- 
The energy $E_{\rm AA}$ of the pair of atoms is given in Eq.~\eqref{E-AA}.

%%%%%%%%%%%%%%%%%%%%%%%%%%%%%%%%%%%%%%%%%%%%%%%
\subsection{Matrix Elements}
%%%%%%%%%%%%%%%%%%%%%%%%%%%%%%%%%%%%%%%%%%%%%%%

Because the system is homogeneous, 
the analog of Eq.~\eqref{homo-AA} can be used to express the 
contact operators $C$ on the right side of Eq.~\eqref{comp-final-state-atom} 
in terms of contact density operators at the same position $\bm{r}$.
The wave-vector delta function in Eq.~\eqref{homo-AA} 
reduces to $\delta^3(\bm{K}-\bm{k}_{\rm D})$, 
and it can be used to integrate over $\bm{K}$. 
The frequency delta function in Eq.~\eqref{Gamma1} reduces to  
%-----------------
\begin{equation}
 \sum_{\pm}2\pi\delta\big((E_{\rm AA}-E_{\rm D})/ \hbar \pm \omega\big)= 
 2\pi\delta \big(\omega - \hbar/m \bar{a}^2  - \hbar k^2/m \big).
\label{delta-omega:minus}
\end{equation}
%-----------------
In the sum over $\pm\omega$, only the $-\omega$ term contributes. 

The expression for the transition rate has been reduced to matrix elements 
of the contact density operator of the form 
$\langle {\bm{K}}, {\bm{k}}|{\cal C}({\bm r})|{\bm k}_{\rm D}\rangle$. 
The matrix element is the complex conjugate of Eq.~\eqref{<D|C|AA>}.
The Gaussian integral over $\bm{K}$ can be evaluated analytically. 
The sum over final states of the matrix element in Eq.~\eqref{comp-final-state-atom}
 reduces to
%----------------- 
\begin{equation}
\sum_{f} \big| \langle f |  C\rho_{\rm gas}C | f \rangle \big|=
64\pi N_{\rm D}\int_{0}^{\infty} dk\hspace{0.1 cm} \frac{k^2\bar{a}}{1+k^2\bar{a}^2}
\hspace{0.1 cm}.
\label{final-contact}
\end{equation}
%----------------- 
Before integrating over $k$,
this must be multiplied by the frequency delta function in Eq.~\eqref{delta-omega:minus}.

%%%%%%%%%%%%%%%%%%%%%%%%%%%%%%%%%%%%%%%%%%%%%%%
\subsection{Harmonic Dissociation Rate}
%%%%%%%%%%%%%%%%%%%%%%%%%%%%%%%%%%%%%%%%%%%%%%%

Our final result for the harmonic disssociation rate $\Gamma_{1}^{(1)}(\omega)$ 
in the homogeneous gas can be obtained 
from Eq.~\eqref{Gamma1} by replacing $\sum_f |\langle f|C|i\rangle|^2$
by the right side of Eq.~\eqref{final-contact},
replacing the sum of frequency delta functions
by the right side of Eq.~\eqref{delta-omega:minus}, 
and  then using the delta function to integrate over $k$.
The local density approximation can be implemented by making the substitution 
for $N_{\rm D}$ in  Eq.~\eqref{LDA-D}. 
The threshold angular frequency for dissociation is $\hbar/m \bar a^2$:
the absorption of smaller energy is not enough to break up the dimer. 
For $\omega > \hbar/m \bar a^2$, the harmonic dissociation rate is 
%----------------- 
\begin{equation}
\Gamma_{1}^{(1)}(\omega)  = 
\frac{\hbar^2 }
{m^2 a_{\rm bg}^2 \bar{a}}\left(\frac{b\Delta}{(\Delta + B_0 - \bar B)^2}\right)^2
\left( \int \! d^3r \, n_{\rm D}(\bm{r}) \right)\\
\frac{(m \omega/\hbar - 1/\bar a^2)^{1/2}}{\omega}.
\label{Gamma-thermal-dimer} 
\end{equation}
%----------------- 
(If the universal dimers are composed of identical bosons, 
the harmonic dissociation rate is given by this same expression.) 
The harmonic dissociation rate in Eq.~\eqref{Gamma-thermal-dimer}
has a maximum at $\omega =  2(\hbar/m \bar a^2)$,
which is twice the threshold angular frequency.
For large frequency, the rate decreases very slowly as $\omega^{-1/2}$.
The dissociation rate is independent of the temperature $T$. 
This may be surprising at first, 
but it is related to the fact that the contact 
of a thermal gas of dimers is independent of $T$.

%%%%%%%%%%%%%%%%%%%%%%%%%%%%%%%%%%%%%%%%%%%%%%%
\subsection{First-Order Subharmonic Dissociation Rate}
%%%%%%%%%%%%%%%%%%%%%%%%%%%%%%%%%%%%%%%%%%%%%%%

According to Eq.~\eqref{Gamma2-2:simple}, 
the contribution $\Gamma_{2}^{(2)}(\omega)$ to the subharmonic dissociation rate 
from the first-order perturbation in $H_{\rm 2}$
can be expressed in terms of the harmonic dissociation rate 
in Eq.~\eqref{Gamma-thermal-dimer} at twice the frequency.
The threshold angular frequency for subharmonic dissociation 
is $\frac12(\hbar/m \bar a^2)$.
For $\omega > \frac12(\hbar/m \bar a^2)$, the transition rate is 
%----------------- 
\begin{equation}
\Gamma_{2}^{(2)}(\omega)  = 
\frac{\hbar^2 }
{8m^2 a_{\rm bg}^2 \bar{a}}\left(\frac{b^2\Delta}{(\Delta + B_0 - \bar B)^3}\right)^2
\left( \int \! d^3r \, n_{\rm D}(\bm{r}) \right)\\
\frac{(2m \omega/\hbar - 1/\bar a^2)^{1/2}}{\omega}.
\label{sub-transition-disso} 
\end{equation}
%----------------- 
This contribution to the subharmonic dissociation rate has a maximum at 
$\omega =  \hbar/m \bar a^2$, which is twice the threshold angular frequency.
The height of the peak is smaller than that of the harmonic dissociation rate
by a factor of $[b/(\Delta + B_0 - \bar B)]^2/4$.

%\newpage

%%%%%%%%%%%%%%%%%%%%%%%%%%%%%%%%%%%%%%%%%%%%%%%%
%
%       SubHarmonic Transition
%
%%%%%%%%%%%%%%%%%%%%%%%%%%%%%%%%%%%%%%%%%%%%%%%%%

\section{Subharmonic Association rate}
\label{sec:subharmassoc}

In this section, we calculate the subharmonic association rate 
in a thermal gas of atoms from the second-order
perturbation in $H_{1}$. 
We will find that this contribution is much larger than
that from the first-order perturbation in $H_{2}$
if $\bar B$ is near a Feshbach resonance.

%%%%%%%%%%%%%%%%%%%%%%%%%%%%%%%%%%%%%%%%%%%%%%%
\subsection{Initial, Final, and Intermediate States}
%%%%%%%%%%%%%%%%%%%%%%%%%%%%%%%%%%%%%%%%%%%%%%%

We first consider a homogeneous gas consisting of $ N_1$ atoms of spin state 1 
and $N_2$ atoms of spin state 2 in a volume $V$. 
If the gas is in thermal equilibrium, 
the harmonic transition rate is given by Eq.~\eqref{Gamma2-11}
with the substitution in Eq.~\eqref{|i><i|-rho},
where $\rho=\rho_{\rm gas}$ is the density matrix for the thermal gas of atoms.
In the low-density limit where 3-body and higher-body correlations can be neglected, 
the density matrix $\rho_{\rm gas}$
can be expressed in terms of the density matrix $\rho_{\rm pair}$ 
for a pair of atoms in thermal equilibrium, as in Eq.~\eqref{density-matrix-gas}.
That density matrix $\rho_{\rm pair}$ is given in
Eq.~\eqref{rho-pair}.  The sum over final states reduces to an integral over the 
wave vector $\bm{k}_{\rm D}$ of the dimer, as in Eq.~\eqref{comp-final-state-therm}.

Once the matrix elements have been reduced to matrix elements 
in the two-atom sector, the sum over intermediate states 
in Eq.~\eqref{Gamma2-11} reduces to a sum over atom-pair states and dimer states.
If the initial state is an atom pair with total energy $E_{\rm AA}$,
the sum over states is
%-----------------
\begin{equation}
 \sum_{m}\frac{ |m\rangle\langle m|}{\omega_{mi}\pm\omega} =
 \int_{\bm{K}'}\int_{\bm{k}'}
 \frac{|{\bm{K}'}, {\bm{k}'}\rangle\langle {\bm{k}'}, {\bm{K}'}|}{(E_{\rm AA}'-E_{\rm AA})/\hbar \pm\omega}
 + \int_{\bm{k}'_{\rm D}}
 \frac{|{\bm k}'_{\rm D}\rangle\langle {\bm k}'_{\rm D}|}{(E_{\rm D}'-E_{\rm AA})/\hbar \pm\omega},
 \label{intermediate-state}
\end{equation}
%-----------------
where $E_{\rm AA}'$ and $E_{\rm D}'$ are given by 
Eqs.~\eqref{E-AA} and \eqref{E-D} with primes on the wavenumber variables.
In the transition rate given by inserting Eq.~\eqref{homo-D} into Eq.~\eqref{Gamma2-11},
there are four  possibilities for the intermediate states
$| m \rangle$ and $| m' \rangle$ in the amplitude and its complex conjugate: 
each one can be either an atom pair or a dimer.
The transition rate can be expressed accordingly as the sum of four terms:
%-----------------
\begin{equation}
 \Gamma_{2}^{(1,1)}=
 \Gamma_{\rm AA,AA}+\Gamma_{\rm D,D}+ \Gamma_{\rm AA,D}+ \Gamma_{\rm D,AA}.
 \label{total-transition}
\end{equation}
%-----------------
We will calculate each of these terms individually.

%%%%%%%%%%%%%%%%%%%%%%%%%%%%%%%%%%%%%%%%%%%%%%%
\subsection{Matrix Elements }
\label{Asso-matrix}
%%%%%%%%%%%%%%%%%%%%%%%%%%%%%%%%%%%%%%%%%%%%%%%

Because the system is homogeneous, the matrix elements of the contact $C$
in Eq.~\eqref{Gamma2-11} can be expressed in terms of matrix elements 
of the contact density operator using Eq.~\eqref{homo-D}.
In addition to the matrix element in Eq.~\eqref{<D|C|AA>} and its complex conjugate, 
we also need the matrix elements of $ {\cal C}(\bm r)$ between
atom-pair states and between dimer states. 
They are calculated in 
Appendix~\ref{appendix-2} and given in Eqs.~\eqref{matrix-element-2-1}
and Eqs.~\eqref{matrix-dimer-3}:
\begin{subequations}
\begin{eqnarray}
\langle \bm{K}', \bm{k}' | {\cal C}(\bm r) |\bm{K}, \bm{k}\rangle
&=&\frac{16\pi^2 \bar a^2}{(1+ i \bar a k')(1- i \bar a k)},
\label{<AA|C|AA>}
\\
\langle \bm{k}_{\rm D}' | {\cal C}(\bm r) |\bm{k}_{\rm D}\rangle
&=&8\pi/\bar a.
\end{eqnarray}
\end{subequations}

The integrals over the total wave vectors of the intermediate states
and over the wave vector of the final-state dimer can be evaluated using 
the delta functions in Eq.~\eqref{homo-D}.
The Gaussian integral over the total wave vector of the initial atom-pair state
can then  be evaluated analytically.
The frequency delta function reduces to
%-----------------
\begin{equation}
 \sum_{\pm}2\pi\delta\big( (E_{\rm D} - E_{\rm AA})/\hbar \pm 2\omega\big)= 
 2\pi\delta \big(2\omega - \hbar/m \bar{a}^2  - \hbar k^2/m \big).
\label{delta-omega:plus2}
\end{equation}
%-----------------
In the sum over $\pm2\omega$, only the $+2\omega$  term contributes. 

%%%%%%%%%%%%%%%%%%%%%%%%%%%%%%%%%%%%%%%%%%%%%%%
\subsubsection{Intermediate atom-pair states }
\label{Asso-atompair}
%%%%%%%%%%%%%%%%%%%%%%%%%%%%%%%%%%%%%%%%%%%%%%%

The contribution from intermediate atom-pair states to the factor in the
transition rate involving matrix elements reduces to
\begin{eqnarray}
\sum_i \sum_f  \left| \sum_{m\ne i} 
\frac{\langle f | C | m \rangle\langle m |C|i\rangle}{\omega_{mi}+\omega} \right|^2
N_1 N_2\langle i |\rho_{\rm pair}| i \rangle
&=&
8192\sqrt{2}\pi\frac{m^2\lambda_{\rm T}^3}{\hbar^2\bar{a}} \frac{N_1N_2}{V}
\nonumber\\
&& \hspace{-6cm}\times 
\int_0^{\infty} \!\!\!\! dk
\frac{k^2 \exp (-\beta\hbar^2k^2/m )}{k^{2}+1/\bar{a}^2}
\left[\int_0^{\infty} \!\!\!\! dk^{'}
\frac{k^{'2}}{(k^{'2}+1/\bar{a}^2)(k^{'2}-k^2 + m\omega/\hbar)}\right]^2.
\label{subhar-AA,AA}
\end{eqnarray}
Before integrating over $k$, this must be multiplied by the frequency delta function
in Eq.~\eqref{delta-omega:plus2}.

The threshold angular frequency for subharmonic association is 
$ \frac12(\hbar/m \bar{a}^2)$.
For $\omega > \hbar/m \bar{a}^2$, the integral over $k'$ in 
Eq.~\eqref{subhar-AA,AA} has a pole on the integration contour.
In this region of $\omega$, this contribution to the transition rate is 
a subleading correction of order $b^4$ to the harmonic transition rate of order
$b^2$ in Eq.~\eqref{Gamma-thermal}.
We therefore consider only  the 
frequency interval $\frac12(\hbar/m \bar{a}^2) < \omega < \hbar/m \bar{a}^2$,
where this contribution  to the transition rate  is leading order in $b$.
In this region of $\omega$, the integral over $k^{'}$ 
in Eq.~\eqref{subhar-AA,AA} is
\begin{equation}
\int_0^{\infty}  \!\!\!\!dk^{'}
\frac{k^{'2}}{(k^{'2}+1/\bar{a}^2)(k^{'2}-k^2 + m\omega/\hbar)}
=\frac{\pi\bar{a}}{2 \big(1+\bar{a}\sqrt{m\omega/\hbar-k^2}\,\big)}.
\label{contour}
\end{equation}
The frequency  delta function in Eq.~\eqref{delta-omega:plus2}
can be used to evaluate the integral over $k$ in 
Eq.~\eqref{subhar-AA,AA}.  The resulting contribution to the transition rate is
\begin{equation}
\Gamma_{\rm{AA,AA}}  = 
\frac{\sqrt{2}\hbar^2 \lambda_{T}^3\bar{a}}
       {4 m^2 a_{\rm bg}^4 }\left(\frac{b\Delta}{(\Delta+B_0-\bar{B})^2}\right)^4
\frac{N_1N_2}{V}
\frac{\kappa(2\omega)\exp\big(-\beta\hbar^2\kappa^2(2\omega)/m\big)}
       {\omega\big(1+\sqrt{1-m\omega \bar{a}^2/\hbar}\, \big)^2 },
    \label{Gamma-subharmonic-atom-atom}
\end{equation}
where $\kappa(2\omega)$ is given in Eq.~\eqref{two-omega-k}.

%%%%%%%%%%%%%%%%%%%%%%%%%%%%%%%%%%%%%%%%%%%%%%%
\subsubsection{Intermediate dimer states}
\label{Asso-dimer}
%%%%%%%%%%%%%%%%%%%%%%%%%%%%%%%%%%%%%%%%%%%%%%%

The contribution from  intermediate dimer states to the factor in the
transition rate involving matrix elements reduces to
\begin{eqnarray}
&&  \sum_i \sum_f \left| \sum_{m} 
\frac{\langle f | C | m \rangle\langle m |C|i\rangle}{\omega_{mi}+\omega} \right|^2
N_1 N_2 \langle i |\rho_{\rm pair}| i \rangle
\nonumber\\
&& \hspace{2cm}=
8192\sqrt{2}\pi^3\frac{m^2 \lambda_{\rm T}^3}{\hbar^2 \bar{a}^3}\frac{N_1N_2}{V}
\int_0^\infty \!\!\!\!dk
\frac{k^2\exp (-\beta\hbar^2k^2/m )}{(k^{2}+1/\bar{a}^2)(k^2+ 1/\bar{a}^2- m\omega/\hbar)^2}.
\label{subhar-D,D}
\end{eqnarray}
Before integrating over $k$, this must be multiplied by the frequency delta function
in  Eq.~\eqref{delta-omega:plus2}.
The frequency  delta function can be used to evaluate the integral over $k$ in 
Eq.~\eqref{subhar-D,D}.  The resulting contribution to the transition rate is
\begin{equation}
\Gamma_{\rm D,D}  = 
\frac{\sqrt{2}\hbar^4 \lambda_{T}^3}
    { m^4 a_{\rm bg}^4 \bar{a}^3}
\left(\frac{b\Delta}{(\Delta+B_0-\bar{B})^2}\right)^4
\frac{N_1N_2}{V}
\frac{\kappa(2\omega)}{\omega^3}
    \exp\left(-\beta\hbar^2\kappa^2(2\omega)/m\right).
    \label{Gamma-subharmonic-dimer-dimer} 
\end{equation}
%

%%%%%%%%%%%%%%%%%%%%%%%%%%%%%%%%%%%%%%%%%%%%%%%
\subsubsection{Interference between Atom-Pair and Dimer States}
\label{Asso-cross}
%%%%%%%%%%%%%%%%%%%%%%%%%%%%%%%%%%%%%%%%%%%%%%%

The contribution to the factor in the
transition rate involving matrix elements  from  intermediate atom-pair states  
in the amplitude and from  intermediate dimer states in its complex conjugate reduces to
\begin{eqnarray}
&& \sum_i  \sum_f
\sum_{m} 
\frac{\langle f | C | m \rangle\langle m |C|i\rangle}{\omega_{mi}+\omega} 
\sum_{m'} 
\frac{\langle i | C | m' \rangle\langle m' |C|f\rangle}{\omega_{m'i}+\omega} 
N_1 N_2 \langle i |\rho_{\rm pair}| i \rangle
\nonumber \\
&&  \hspace{1cm}=
-8192\sqrt{2}\pi^2\frac{m^2\lambda_{\rm T}^2}{\hbar^2\bar{a}^2}\frac{ N_1N_2}{V}
\int_0^\infty \!\!\!\!dk
\frac{k^2\exp (-\beta\hbar^2k^2/m )}{(k^{2}+1/\bar{a}^2)(k^2+1/\bar{a}^2 - \omega/\hbar)}
\nonumber\\
&& \hspace{5cm}\times
\int_0^{\infty}\!\!\! dk^{'}
\frac{k^{'2}}{(k^{'2}+1/\bar{a}^2)(k^{'2}-k^2 + m\omega/\hbar)}.
\label{subhar-AA,D}
\end{eqnarray}
Before integrating over $k$, this must be multiplied by the frequency delta function
in  Eq.~\eqref{delta-omega:plus2}.
The integral over $k^{'}$ is given in Eq.~\eqref{contour}.
The frequency  delta function can be used to evaluate the integral over $k$ in 
Eq.~\eqref{subhar-D,D}.  The resulting contribution to the transition rate is
\begin{equation}
\Gamma_{\rm {AA,D}}  =-
\frac{\hbar^3 \lambda_{T}^3}
    {\sqrt{2} m^3 a_{\rm bg}^4  \bar{a}}
\left(\frac{b\Delta}{(\Delta+B_0-\bar{B})^2}\right)^4
\frac{}{}
\frac{N_1N_2}{V}
\frac{\kappa(2\omega)\exp\big(-\beta\hbar^2\kappa^2(2\omega)/m\big)}
       {\omega^2 \big(1+\sqrt{1-m\omega\bar{a}^2/\hbar}\, \big)}.
 \label{Gamma-subharmonic-atom-dimer} 
\end{equation}
The contribution $\Gamma_{\rm D,AA}$ from  intermediate dimer states  
in the amplitude and intermediate atom-pair states in its complex conjugate 
is the same as Eq.~\eqref{Gamma-subharmonic-atom-dimer}
for frequencies in the range
$\frac12 (\hbar/m \bar a^2) < \omega < \hbar/m \bar a^2$.
The contributions $\Gamma_{\rm AA,D}=\Gamma_{\rm D,AA}$ are  negative,
because there is destructive interference 
between atom-pair and dimer intermediate states.

%%%%%%%%%%%%%%%%%%%%%%%%%%%%%%%%%%%%%%%%%%%%%%%
\subsection{Total subharmonic transition rate}
\label{Asso-total}
%%%%%%%%%%%%%%%%%%%%%%%%%%%%%%%%%%%%%%%%%%%%%%%

The total subharmonic transition rate in Eq.~\eqref{total-transition} 
from the second-order perturbation in $H_{1}$ is given by adding 
Eqs.~\eqref{Gamma-subharmonic-atom-atom} and
\eqref{Gamma-subharmonic-dimer-dimer} and 
twice Eq.~\eqref{Gamma-subharmonic-atom-dimer}.
The local density approximation can be implemented by making the substitution 
for $N_1 N_2/V$ in  Eq.~\eqref{LDA-AA}. 
For frequencies in the range $\frac12 (\hbar/m \bar a^2) < \omega < \hbar/m \bar a^2$,
the subharmonic transition rate is
\begin{eqnarray}
 \Gamma_2^{(1,1)}(\omega)&=&
 \frac{\sqrt{2}\hbar^2\bar{a}}{4m^2a_{\rm bg}^4}
 \left(\frac{b\Delta}{(\Delta+B_0-\bar{B})^2}\right)^4
 \left(\int d^3r\hspace{0.2 cm} n_1({\bm r})n_2({\bm r})\right)
 \nonumber\\
 &&\hspace{0.5cm}\times
 \frac{\lambda_{T}^3\kappa(2\omega)}{\omega}
 \exp\left(-\beta\hbar^2\kappa^2(2\omega)/m\right)
 \left(\frac{1}{1+\sqrt{1-m\omega\bar{a}^2/\hbar}}-\frac{2}{m\omega\bar{a}^2/\hbar}\right)^2,
 \label{total-rate}
\end{eqnarray}
where $\kappa(2\omega)$ is given by Eq.~\eqref{two-omega-k}.
(The corresponding result for identical bosons can be obtained by replacing 
$n_1({\bm r})n_2({\bm r})$ by $n^2({\bm r})/2$, where $n({\bm r})$ 
is the local  number density.)

If $k_BT \ll \hbar^2/m \bar a^2$, the subharmonic association  rate in Eq.~\eqref{total-rate}
has a narrow peak
with a maximum when $\omega$ is above 
the threshold $\frac12(\hbar/m \bar a^2)$ by approximately $k_BT/4\hbar$.
In the region near the threshold and the peak,
the largest contribution comes from intermediate dimer states. 
The intermediate atom-pair states give a contribution
that is smaller  at threshold by a factor of $(3-2\sqrt2)/8 \approx 0.021$. 
The cross terms give a negative contribution that is smaller 
at threshold by a factor of $(2-\sqrt2)/2 \approx 0.29$.
The subharmonic association rate in Eq.~\eqref{total-rate}
is much smaller than the harmonic association rate in  
Eq.~\eqref{Gamma-thermal}.
The ratio of their maximum values is 
\begin{equation}
\frac{ \Gamma_{2,{\rm max}}^{(1,1)}}{ \Gamma_{1,{\rm max}}^{(1)}}=
2.914  \left(\frac{b\Delta}{(\Delta+B_0-\bar{B})^2}\right)^2
  \left( \frac{\bar a}{a_{\rm bg}} \right)^2
  \left( 1 - 1.21 \frac{k_BT m\bar a^2}{\hbar^2} + \ldots\right).
 \label{ratio-assoc}
\end{equation}

The contribution $\Gamma_2^{(2)}$
to the subharmonic transition rate from the first-order perturbation 
in $H_2$ is given in Eq.~\eqref{sub-transition-asso}.
Near the subharmonic threshold frequency, $\Gamma_2^{(1,1)}$
differs from $\Gamma_2^{(2)}$ 
by a factor of $11.6\, (\bar a/a_{\rm bg})^2[\Delta/(\Delta+B_0-\bar{B})]^2$.  
If $\bar B$ is near the Feshbach resonance, 
$\Gamma_2^{(2)}$ is much smaller.
It is therefore unnecessary to consider interference between
the first-order perturbation in $H_1$ and the second-order perturbation in $H_2$.

%\newpage

%%%%%%%%%%%%%%%%%%%%%%%%%%%%%%%%%%%%%%%%%%%%%%%
\section{Subharmonic Dissociation rate}
\label{sec:subharmdissoc}
%%%%%%%%%%%%%%%%%%%%%%%%%%%%%%%%%%%%%%%%%%%%%%%

In this section, we calculate the subharmonic dissociation rate 
in a thermal gas of dimers from the second-order
perturbation in $H_{1}$. 
We first consider a homogeneous gas of $ N_{\rm D}$ dimers in a volume $V$
in thermal equilibrium at temperature $T$.
The subharmonic transition rate is given by Eq.~\eqref{Gamma2-11}
with Eq.~\eqref{homo-D} inserted and with $|i\rangle \langle i|$ replaced
by the density matrix $\rho_{\rm gas}$ for the thermal gas of dimers. 
In the low-density limit where correlations between dimers can be neglected, 
$\rho_{\rm gas}$ can be expressed in terms of the density matrix $\rho_{\rm dimer}$ 
for a dimer in thermal equilibrium, as in Eq.~\eqref{density-matrix-dimergas}. 
The density matrix $\rho_{\rm dimer}$ is given in
Eq.~\eqref{rho-dimer}. The sum over final states reduces to an integral 
over center-of-mass wave vector $K$ and relative wave vector $k$ 
of a pair of atoms, as in Eq.~\eqref{comp-final-state-atom}. 
The frequency delta function  reduces to
\begin{equation}
 \sum_{\pm}2\pi\delta\big( (E_{\rm AA} - E_{\rm D})/\hbar \pm 2\omega\big)= 
 2\pi\delta \big(2\omega - \hbar/m \bar{a}^2  - \hbar k^2/m \big).
\label{delta-omega:plus3} 
\end{equation}
In the sum over $\pm2\omega$, only the $-2\omega$  term contributes.
In the sums over intermediate states in the amplitude and its complex conjugate,
both intermediate states can be either an atom pair or a dimer.
The calculation of the individual contributions
proceeds in the same way as for the association rate.
They can be 
obtained from those in Eqs.~\eqref{Gamma-subharmonic-atom-atom}, 
\eqref{Gamma-subharmonic-dimer-dimer}, and \eqref{Gamma-subharmonic-atom-dimer} 
by replacing $N_1N_2/V$ by $N_{\rm D}$ and replacing 
$\lambda_{\rm T}^3\exp\left(-\beta\hbar^2\kappa^2(2\omega)/m\right)$ by $\sqrt{2}/4$.

In the local density approximation,
the factor $N_{\rm D}$ is replaced by $\int d^3r\, n_{\rm D}(\bm r)$.
For frequencies in the range $\frac12 (\hbar/m \bar a^2) < \omega < \hbar/m \bar a^2$,
the subharmonic dissociation rate is
\begin{eqnarray}
 \Gamma_{2}^{(1,1)}(\omega)&=&
 \frac{\hbar^2\bar{a}}{8m^2a_{\rm bg}^4}
 \left(\frac{b\Delta}{(\Delta+B_0-\bar{B})^2}\right)^4
 \left(\int d^3r\hspace{0.2 cm} n_{\rm D}({\bm r})\right)
 \nonumber\\
 &&\hspace{1.5 cm}\times
  \frac{\kappa(2\omega)}{\omega}
\left(\frac{1}{1+\sqrt{1-m\omega\bar{a}^2/\hbar}}-\frac{2}{m\omega\bar{a}^2/\hbar}\right)^2,
 \label{total-rate-2}
\end{eqnarray}
where $\kappa(2\omega)$ is given by Eq.~\eqref{two-omega-k}.
(If the universal dimers are composed of identical bosons, 
the subharmonic dissociation rate is given by this same expression.)
Like the harmonic dissociation rate in Eq.~\eqref{Gamma-thermal-dimer},
the subharmonic dissociation rate  in Eq.~\eqref{total-rate-2}
is independent of the temperature $T$.
The subharmonic dissociation rate has a maximum 
at an angular frequency $\omega$ that is above the
threshold $\frac12 (\hbar/m \bar a^2)$ by approximately $0.082(\hbar/m \bar a^2)$.
The subharmonic dissociation rate in Eq.~\eqref{total-rate}
is much smaller than the harmonic dissociation rate in  
Eq.~\eqref{Gamma-thermal-dimer}.
The ratio of their maximum values is 
\begin{equation}
\frac{ \Gamma_{2,{\rm max}}^{(1,1)}}{ \Gamma_{1,{\rm max}}^{(1)}}=
1.39  \left(\frac{b\Delta}{(\Delta+B_0-\bar{B})^2}\right)^2
  \left( \frac{\bar a}{a_{\rm bg}} \right)^2.
 \label{ratio-dissoc}
\end{equation}

The contribution $\Gamma_2^{(2)}$
to the subharmonic transition rate from the first-order perturbation 
in $H_2$ is given in Eq.~\eqref{sub-transition-disso}.
Near the subharmonic threshold frequency, $\Gamma_2^{(1,1)}$
differs from $\Gamma_2^{(2)}$ 
by a factor of $11.6\, (\bar a/a_{\rm bg})^2[\Delta/(\Delta+B_0-\bar{B})]^2$.  
If $\bar B$ is near the Feshbach resonance, 
$\Gamma_2^{(2)}$ is much smaller.
It is therefore unnecessary to consider interference between
the first-order perturbation in $H_1$ and the second-order perturbation in $H_2$.

%\newpage

%%%%%%%%%%%%%%%%%%%%%%%%%%%%%%%%%%%%%%%%%%%%%%%
\section{Application to $\bm{^7}$Li atoms}
\label{sec:7Li}
%%%%%%%%%%%%%%%%%%%%%%%%%%%%%%%%%%%%%%%%%%%%%%%

An experiment on the association of $^7{\rm Li}$ atoms 
into universal dimers using a modulated magnetic field
was performed at Rice University 
by Dyke, Pollack, and  Hulet \cite{Hulet1302}. 
The $^7\rm{Li}$ atoms were in $|F=1,m_F=1\rangle$ state.
The scattering length was controlled using the Feshbach resonance 
at $B_{0}=737.7$~G.
The other parameters in the expression for the scattering length in Eq.~\eqref{a-B},
as determined in  Ref.~\cite{Hulet1302},
are $a_{\rm bg}=-20.0\hspace{0.1 cm}a_0$ and $\Delta=-174$~G.
The bias magnetic field was set to $\bar B =734.5$~G. 
The measurements were carried out for both a Bose-Einstein condensate
and a thermal gas of  $^7{\rm Li}$ atoms.
Magneto-association into dimers was observed through the loss of atoms,
presumably from inelastic collisions of the dimers with atoms.

In the experiment on the BEC of $^7$Li atoms in Ref.~\cite{Hulet1302}, 
they observed a narrow loss resonance as a function of the frequency,
with the fraction of atoms remaining decreasing almost to 0.
The position of this resonance was used to measure the binding energy 
of the dimer to be $h$(450~kHz), 
corresponding to $\bar a \approx1100~a_0$.

The experiment on the thermal gas of  $^7{\rm Li}$ atoms 
in Ref.~\cite{Hulet1302} was carried out
at three combinations of the amplitude $b$ of the modulated magnetic field
and the temperature $T$: $(b,T)$=$(0.57\,{\rm G},3\, {\rm \mu K})$, 
$(0.14\, {\rm G},3\, {\rm \mu K})$, and $(0.57\, {\rm G},10\, {\rm \mu K})$. 
The duration of modulation was in the range from 25~$\mu$s to 500~$\mu$s,
but its value was not specified for each individual set $(b,T)$.
The fraction of atoms remaining after the modulation time
was measured as a function of frequency of the oscillating field.
The data were fit to convolutions of Lorentzians 
with the thermal Boltzmann distributions.
For each set $(b,T)$, there was a harmonic peak just above $\omega_D$.
A subharmonic peak just above $\omega_D/2$ was
evident only for $(0.57\, {\rm G},\, 3{\rm \mu K})$. 
For these values of $(b,T)$, the minimum fractions remaining 
were about 0.3 in the harmonic peak
and about 0.5 in the subharmonic peak.

%%%%%%%%%%%%%%%%%%%%%%%%%%%%%%%%%%
\begin{figure}[tbh]
 \centering
 \includegraphics[angle=270,scale=0.6]{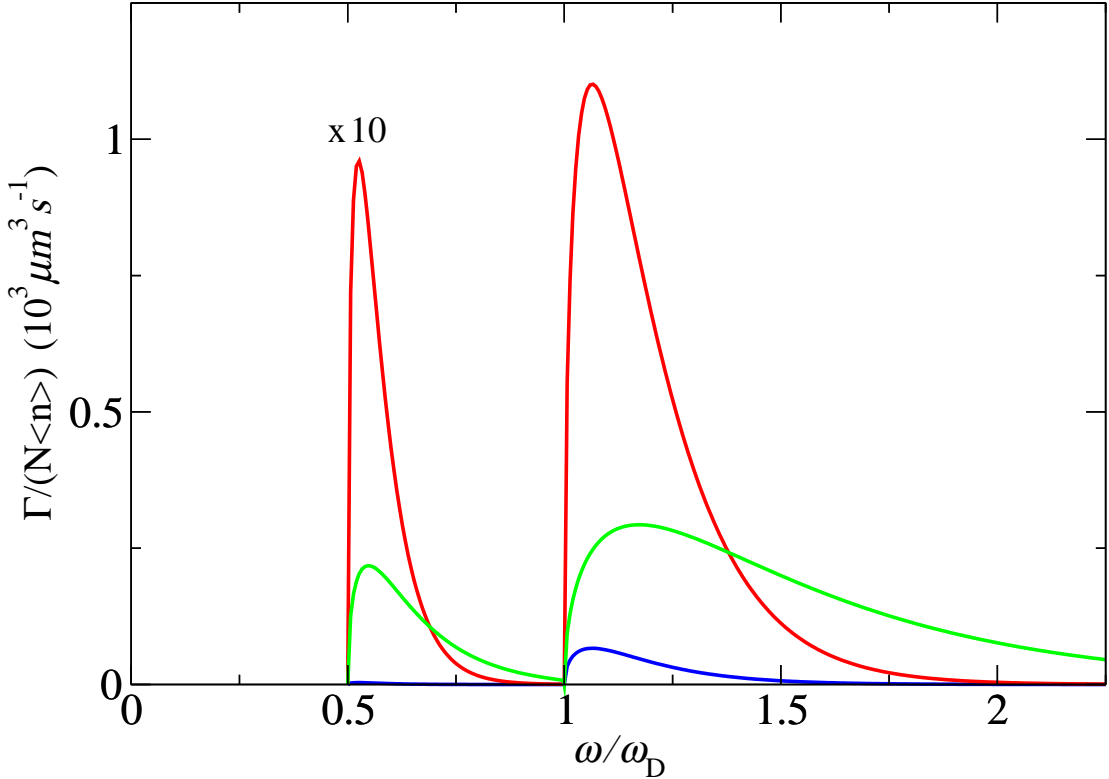}
%\includegraphics[angle=270,scale=0.6]{Figures/assoc.pdf}
 % subharmonic-plot.pdf: 0x0 pixel, 0dpi, nanxnan cm, bb=
 \caption{
Association rate $\Gamma/N\langle n \rangle$ 
 for a thermal gas of $^7{\rm Li}$ atoms 
 as a function of the modulation frequency of the magnetic field.
The angular frequency $\omega$ is normalized to the angular binding frequency
$\omega_D$ of the dimer.   
The bias field is $\bar B =734.5$~G, which corresponds to $\bar a  \approx 1100~a_0$.
The three sets of curves are for different values of the modulation amplitude 
$b$ and the temperature $T$:
 $(b,T)$=$(0.57\,{\rm G},3\, {\rm \mu K})$ (highest peak value, red), 
 $(0.57\, {\rm G},10\, {\rm \mu K})$ (green),
 and $(0.14\, {\rm G},3\, {\rm \mu K})$ (lowest peak value, blue). 
The association rates in the 
region $\omega < \omega_D$ have been multiplied by 10 to make the
subharmonic transitions more visible.
}
 \label{figure-assoc}
\end{figure}
%%%%%%%%%%%%%%%%%%%%%%%%%%%%%%%%%%

Since the number of atoms in the thermal clouds and 
the modulation times for each set of $(b,T)$ were not specified in Ref.~\cite{Hulet1302},
we are unable to make quantitative comparisons with our theoretical results.
In Fig.~\ref{figure-assoc}, we show our results for the association rates 
as functions of the angular frequency $\omega$ for the three 
sets of values of $(b,T)$ for which the atom loss was measured in Ref.~\cite{Hulet1302}.
The ratio of $k_BT/\hbar$ to the binding frequency $\omega_D =\hbar/m \bar a^2$
of the dimer is $0.15$ and $0.49$ at the temperatures $3\,{\rm \mu K}$
and $10\,{\rm \mu K}$, respectively.
Thus the condition $k_BT \ll\hbar^2/m \bar a^2$
is much better satisfied at $3\,{\rm \mu K}$.
In Fig.~\ref{figure-assoc}, the curves for $\omega > \omega_D$ 
are the harmonic association rates 
$\Gamma_1^{(1)}(\omega)$ in Eq.~\eqref{Gamma-thermal}
with $\int d^3r\, n_1 n_2$ replaced by $\int d^3r\, n^2/2$.
The curves for $\omega < \omega_D$ are the subharmonic association rates 
$\Gamma_2^{(1,1)}(\omega)$ in Eq.~\eqref{total-rate}
with $\int d^3r\, n_1 n_2$ replaced by $\int d^3r\, n^2/2$.
The subharmonic association rates $\Gamma_2^{(2)}(\omega)$ are completely negligible:
their peak values are smaller than those for $\Gamma_2^{(1,1)}$
by factors of about $3\times 10^{-5}$. 
The association rates in Fig.~\ref{figure-assoc} are divided by 
$\int d^3r \, n^2(\bm{r}) = N\langle n \rangle $ to obtain rates 
$\Gamma/N\langle n \rangle$ that do not depend on the number of atoms.
The angular frequency is normalized to the angular binding frequency
$\omega_D = \hbar/m \bar a^2$ of the dimer.   
The heights of the harmonic and subharmonic peaks for 
$(b,T)$=$(0.57\,{\rm G},3\, {\rm \mu K})$ are larger than those for 
$(0.14\, {\rm G},3\, {\rm \mu K})$ by the ratio of the values of $b^2$,
which is 12.8.
The maxima of the harmonic association rates are at angular frequencies  $\omega$ 
that are above the threshold $\omega_{\rm D}$ 
by approximately $k_BT/2\hbar$.
The maxima of the subharmonic association rates are at angular frequencies  
that are above the threshold $\frac12 \omega_{\rm D}$ 
by approximately $k_BT/4\hbar$.
The ratios of the maximum of the harmonic peak to the maximum of the 
subharmonic peak are 11.4, 190, and 13.8
for $(b,T)$=$(0.57\,{\rm G},3\, {\rm \mu K})$, $(0.14\, {\rm G},3\, {\rm \mu K})$, 
and $(0.57\, {\rm G},10\, {\rm \mu K})$, respectively.
They can be compared to the ratios 11.9, 198, and 24.1
predicted using Eq.~\eqref{ratio-assoc}.

%%%%%%%%%%%%%%%%%%%%%%%%%%%%%%%%%%
\begin{figure}[tbh]
 \centering
 \includegraphics[angle=270,scale=0.6]{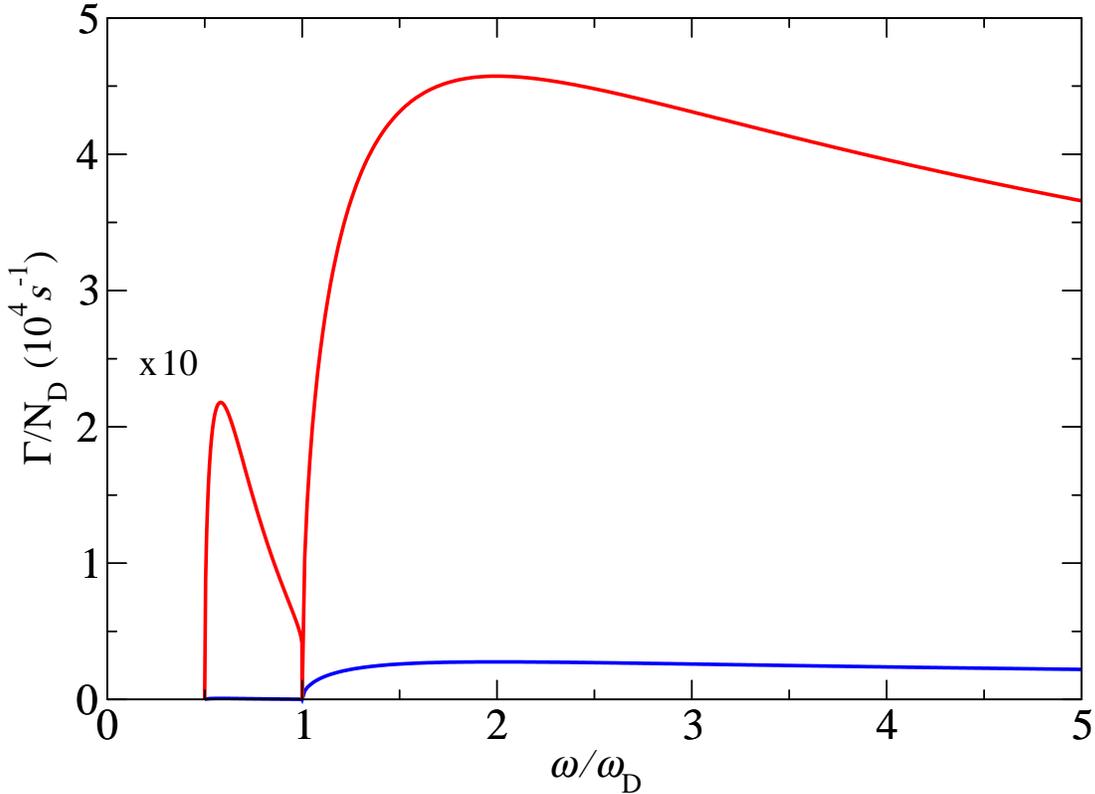}
%\includegraphics[angle=270,scale=0.6]{Figures/dissoc.pdf}
 % subharmonic-plot.pdf: 0x0 pixel, 0dpi, nanxnan cm, bb=
 \caption{
Dissociation rate $\Gamma/N_D$ 
 for a thermal gas of $^7{\rm Li}$ dimers 
 as a function of the angular frequency $\omega$ of the modulated magnetic field.
The angular frequency $\omega$ is normalized to the binding angular frequency
$\omega_D$ of the dimer.   
The bias field is $\bar B =734.5$~G, which corresponds to $\bar a  \approx 1100~a_0$.
The dissociation rate is independent of the temperature.
The two sets of curves are for different values of the modulation amplitude $b$: 
0.57~G (higher peak value, red) and 0.14~G (lower peak value, blue). 
The dissociation rates in the 
region $\omega < \omega_D$ have been multiplied by 10 to make the
subharmonic transitions more visible.
}
 \label{figure-dissoc}
\end{figure}
%%%%%%%%%%%%%%%%%%%%%%%%%%%%%%%%%%

Our results for dissociation rates in a thermal gas of dimers 
as functions of the angular frequency $\omega$ 
are illustrated in Fig.~\ref{figure-assoc} using the two values of $b$ 
for which the atom loss was measured in Ref.~\cite{Hulet1302}.
The dissociation rates in Fig.~\ref{figure-assoc} are divided by 
the number $N_D$ of dimers to obtain rates 
$\Gamma/N_D$ that do not depend on $N_D$.
The angular frequency $\omega$ is normalized to the binding angular frequency
$\omega_D$ of the dimer.   
The curves for $\omega > \omega_D$ are the harmonic dissociation rates 
$\Gamma_1^{(1)}(\omega)$ in Eq.~\eqref{Gamma-thermal-dimer}.
The curves for $\omega < \omega_D$ are the subharmonic dissociation rates 
$\Gamma_2^{(1,1)}(\omega)$ in Eq.~\eqref{total-rate-2}.
The subharmonic association rates $\Gamma_2^{(2)}(\omega)$ are completely negligible:
their peak values are smaller than those for $\Gamma_2^{(1,1)}$
by factors of about $3\times 10^{-5}$.
The heights of the harmonic and subharmonic peaks for 
$b=0.57$~G are larger than those for 
$b=0.14$~G by the ratio of the values of $b^2$, which is 12.8.
The maxima of the
harmonic association rates are at twice the threshold angular frequency  $\hbar/m \bar a^2$.
For large $\omega$, the harmonic rates decrease very slowly as $\omega^{-1/2}$.
The maxima of the subharmonic dissociation rates are above the threshold angular frequency  
$\frac12(\hbar/m \bar a^2)$ by only about 0.08~$\hbar/m \bar a^2$.
The ratio of the maximum of the harmonic peak to the maximum of the 
subharmonic peak is given in Eq.~\eqref{ratio-dissoc}.
The ratios for $b=0.57$~G and $b=0.14$~G
are approximately 21 and 340, respectively.

%%%%%%%%%%%%%%%%%%%%%%%%%%%%%%%%%%%%%%%%
%                                      
%           Discussion                   
%                                      
%%%%%%%%%%%%%%%%%%%%%%%%%%%%%%%%%%%%%%%%

\section{Comparisons and Summary}
\label{sec:discsum}

In this section, we describe previous theoretical treatments 
of the association of atoms into dimers using a modulated magnetic field.
We then summarize our results on association and dissociation
and compare the association results with those from the other approaches.

A theoretical treatment 
of the association of atoms into dimers using an oscillating magnetic field
was presented by Hanna, K\"ohler, and Burnett in 2007 \cite{HKB0609}.   
They used a two-channel model for the two-atom system, with one channel 
consisting of atom pairs interacting through a short-range separable potential 
and a second channel consisting of a single discrete molecular state.
The dimer is an eigenstate of the coupled-channel problem.
The modulation of the magnetic field was taken into account
through the sinusoidal oscillation of the energy of the discrete molecular state.
They solved the time-dependent Schr\"odinger equation 
for the two-channel model numerically to obtain
the probability for association into the dimer as a function of time.
They studied the dependence of the  association probability 
for a homogeneous thermal gas on the frequency and amplitude 
of the oscillating field and on the temperature and density of the gas.
Association into dimers in a Bose-Einstein condensate was
treated in a completely different way by solving numerically an 
integro-differential equation for the mean-field of a homogeneous BEC.
The results for association probabilities in Ref.~\cite{HKB0609}
are all completely numerical.
They can be applied quantitatitively only to homogeneous systems of $^{85}$Rb 
and $^{133}$Cs atoms under the specific conditions considered in the paper.
In order to use their methods to predict the association probability for 
any other conditions, such as different atoms, 
other oscillation parameters, 
different temperature or density for the homogeneous gas, or a trapped gas
with variable density,
it would be necessary to solve their equations numerically
for each set of conditions.
Subharmonic transitions were not considered in Ref.~\cite{HKB0609}.

Brouard and Plata have recently presented a different theoretical treatment 
of the association of atoms into dimers using an oscillating magnetic field \cite{BP1503}.
They also used a two-channel model for the two-atom system, with one channel 
consisting of a continuum of positive-energy atom pair states 
and the second channel consisting of a single discrete molecular state.
The dimer is an eigenstate of the coupled-channel problem.
As in Ref.~\cite{HKB0609}, the modulation of the magnetic field 
was taken into account
through the sinusoidal oscillation of the energy of the discrete molecular state.
However in Ref.~\cite{BP1503}, a time-dependent unitary transformation
was used to move the time-dependence into off-diagonal terms
between the dimer state and the atom-pair states.
They showed that for frequencies near the harmonic resonance, 
the dynamics in the transformed frame can be approximated 
by a time-independent Hamiltonian whose entries are given analytically
as functions of the 
oscillation parameters and the Feshbach resonance parameters.
Subharmonic transitions in an appropriate transformed frame 
are described by a different time-independent Hamiltonian 
whose entries are given analytically.
Some qualitative aspects of the association process were deduced 
from these effective Hamiltonians.
However the results in Ref.~\cite{BP1503} for association probabilities in 
thermal gases of $^{85}$Rb atoms and in Bose-Einstein condensates
of $^{85}$Rb atoms are completely numerical.

In this paper, we have applied the new approach to this problem
that was introduced in Ref.~\cite{LSB:1406}.  It was based on the realization 
that the leading effect of an oscillating magnetic field near a Feshbach resonance 
can be treated as a time-dependent perturbation proportional to the
contact operator $C$.  In Appendix~\ref{appendix-1}, 
we presented a quantum field theory argument 
that the perturbation proportional to $C$ can also be used beyond first order.
Fermi's Golden Rule is used to 
obtain general expressions for transition rates in terms of 
transition matrix elements of $C$.
Our general formula for the harmonic transition rate 
$\Gamma_{1}^{(1)}(\omega)$ in Eq.~\eqref{Gamma1}
comes from the first-order perturbation in $C$ and was
obtained previously in Ref.~\cite{LSB:1406}. 
There is a contribution $\Gamma_{2}^{(2)}(\omega)$ 
to the subharmonic transition rate from the first-order perturbation
in $C$ that is related in a simple way to the harmonic rate 
and is given in Eq.~\eqref{Gamma2-2:simple}.
However the second-order perturbation in $C$ gives another contribution 
$\Gamma_{2}^{(1,1)}(\omega)$ to the subharmonic transition rate 
that  is given in Eq.~\eqref{Gamma2-11}.
Near the Feshbach resonance, $\Gamma_{2}^{(1,1)}$ is larger than 
$\Gamma_{2}^{(2)}$ by a factor of $(\bar a/a_{\rm bg})^2$.

For a homogeneous system,
our general expressions for the transition rates can be simplified
by expressing the transition matrix elements of the contact operator
in terms of transition matrix elements of the contact  density 
operator ${\cal C}$.
The harmonic transition rate 
$\Gamma_{1}^{(1)}$ in Eq.~\eqref{Gamma1}
can be simplified by inserting Eq.~\eqref{homo-AA}.
The subharmonic transition rate 
$\Gamma_{2}^{(1,1)}$ in Eq.~\eqref{Gamma2-11}
can be simplified by inserting Eq.~\eqref{homo-D}.
For a nonhomogeneous system in the local density approximation,
these simplifications can first be used to calculate the
transition rates for the homogeneous system.
Substitutions such as those in Eqs.~\eqref{LDA-AA} and \eqref{LDA-D}
can then be used to obtain the transitions rate for the nonhomogenous system.

To obtain association rates in a thermal gas of atoms
and dissociation rates in a thermal gas of dimers,
we first exploited the low density to reduce 
the  transition matrix elements of ${\cal C}$ in the thermal gas  
to transition matrix elements of ${\cal C}$ in the two-body problem.
Those matrix elements were calculated in Appendix~\ref{appendix-2}
using the quantum field theory formulation of the problem of atoms 
with zero-range interactions.  In the two-atom sector,
this is equivalent to a single-channel model for atoms with large scattering length,
with the dimer arising dynamically as a bound state.
This allowed us to calculate the matrix elements of the contact density analytically.

Our final results for the harmonic and subharmonic association rates 
in a thermal gas of atoms are given in Eqs.~\eqref{Gamma-thermal}
and \eqref{total-rate}.
Our final results for the harmonic and subharmonic disssociation rates 
in a thermal gas of dimers are given in Eqs.~\eqref{Gamma-thermal-dimer}
and \eqref{total-rate-2}.  These results 
are analytic functions of all the relevant parameters:
the oscillation parameters $\omega$, $b$, 
and $\bar a$ or $\bar B$, the Feshbach resonance parameters
$a_{\rm bg}$, $B_0$, and $\Delta$, and the temperature $T$.
The association rates in a thermal gas of fermions with two spin states
depend on the local number densities $n_1(\bm{r})$ and $n_2(\bm{r})$ 
only through the multiplicative factor $\int d^3r\, n_1 n_2$.
The dissociation rates in a thermal gas of dimers
depend on the local number density $n_{\rm D}(\bm{r})$ only through
the multiplicative factor $\int d^3r\, n_{\rm D}$.
Our analytic results should be useful for analyzing experiments 
on association into and dissociation of dimers.
They should also be useful for designing experiments that optimize the 
number of dimers created or destroyed by the modulated magnetic field.
For a thermal gas of atoms with $k_BT \ll \hbar^2/m \bar a^2$, 
the maximum in the harmonic association rate 
is at an angular  frequency $\omega$ that is above 
the threshold $\hbar/m \bar a^2$ by approximately $k_BT/2\hbar$.
The maximum in the subharmonic association rate 
is at an angular  frequency that is above 
the threshold $ \frac12 (\hbar/m \bar a^2)$ by approximately $k_BT/4\hbar$.
For a thermal gas of dimers, 
the maximum in the harmonic dissociation rate 
is at an angular  frequency $\omega$ that is approximately twice 
the threshold $\hbar/m \bar a^2$.
The maximum in the subharmonic dissociation rate 
is at an angular  frequency that is above 
the threshold $ \frac12 (\hbar/m \bar a^2)$ by approximately 
$0.08\, (\hbar/m \bar a^2)$.

Our general results for the harmonic and subharmonic transition rates in terms of 
matrix elements of the contact density operator can also be applied to other systems.
An analytic result for the association rate in a dilute Bose-Einstein condensate 
of identical bosons was given in Ref.~\cite{LSB:1406}. 
It should also be possible to obtain analytic results for superfluids
of fermions with two spin states at zero temperature,
including the dissociation rate of dimers 
in the BEC limit and the dissociation rate of Cooper pairs in the BCS limit.
The dissociation rate of paired fermions in the unitary limit 
is more challenging, but
it is an important  problem because it would allow the first direct 
measurements of the gap for the unitary Fermi gas.
 
\begin{acknowledgments}
This research was supported in part by the
National Science Foundation under grant PHY-1310862
and by the Simons Foundation.
We thank Hudson Smith for valuable discussions.

%\newpage

%%%%%%%%%%%%%%%%%%%%%%%%%%%%%%%%%%%%%%%
%
%         Appendix
%
%%%%%%%%%%%%%%%%%%%%%%%%%%%%%%%%%%%%%

\appendix

%%%%%%%%%%%%%%%%%%%%%%%%%%%%%%%%%%%%%%%%%%%%%
\section{Quantum Field Theory Derivation of Perturbing Hamiltonian}
\label{appendix-1}
%%%%%%%%%%%%%%%%%%%%%%%%%%%%%%%%%%%%%%%%%%%%%

Particles with a scattering length $a$ that is large compared to the range $r_0$ 
of their interactions can be described by a local quantum field theory.
For a fermion with two spin states, there are two fermionic quantum fields 
$\psi_1$ and $\psi_2$.  The interactions of the quantum field theory are made local
by taking the zero-range limit at the expense of introducing an ultraviolet cutoff $\Lambda$
on the momenta of virtual particles.  The interaction Hamiltonian density is
\begin{equation}
{\cal H}_{\rm int} = (\lambda_0/m)\psi_1^{\dagger}\psi_{2}^{\dagger}\psi_2\psi_1,
\label{hamiltonian-density}
\end{equation}
where $\lambda_0$ is the bare coupling constant.
If $\hbar$ is set to 1, $\lambda_0$ has dimensions of length.
The field theory describes particles with scattering length $a$ 
if the bare coupling constant is
\begin{equation}
 \lambda_0=\frac{4\pi}{1/a -2\Lambda/\pi}.
 \label{coupling}
\end{equation}

Matrix elements of the operator $\psi_1^{\dagger}\psi_{2}^{\dagger}\psi_2\psi_1$
diverge as $\Lambda^2$ as the cutoff is increased to $\infty$.  
Since $\lambda_0$ scales as $1/\Lambda$, 
matrix elements of the interaction Hamiltonian density in Eq.~\eqref{hamiltonian-density}
therefore diverge as $\Lambda$.  
In matrix elements of the complete Hamiltonian density, the divergence is cancelled by 
a corresponding divergence in matrix elements of the kinetic energy density.
In matrix elements of the operator $\psi_1^{\dagger}\psi_{2}^{\dagger}\psi_2\psi_1$,
subleading terms that diverge as $\Lambda$ give finite contributions to the energy density.
The contact density operator in the quantum field theory is \cite{Braaten:2008uh}
\begin{equation}
{\cal C} =\lambda_0^2\psi_1^{\dagger}\psi_{2}^{\dagger}\psi_2\psi_1.
\label{contact-density}
\end{equation}
This operator has finite matrix elements, because the divergence
in the matrix element of $\psi_1^{\dagger}\psi_{2}^{\dagger}\psi_2\psi_1$
proportional to $\Lambda^2$
is compensated by a factor of $1/\Lambda^2$ from $\lambda_0^2$.

The local quantum field theory can describe particles with a 
time-dependent scattering length $a(t)$ provided the time scale 
$a/\dot a$ is large compared to the time scale $m r_0^2/\hbar$ 
set by the range.  In the interaction Hamiltonian density
in Eq.~\eqref{hamiltonian-density}, the time-dependent bare 
coupling constant $\lambda_0(t)$ is obtained by replacing $a$ in Eq.~\eqref{coupling}
by $a(t)$.  If the time dependence consists of small deviations
in the inverse scattering length from some value $1/\bar a$,
the bare coupling constant can be expanded 
around the corresponding value $\bar \lambda_0$:
\begin{equation}
\lambda_0(t) = \bar \lambda_0 
- \frac{\bar \lambda_0^2}{4 \pi} \left( \frac{1}{a(t)} - \frac{1}{\bar a} \right) 
+ \frac{\bar \lambda_0^3}{(4 \pi)^2} \left( \frac{1}{a(t)} - \frac{1}{\bar a} \right)^2 + \ldots.
\label{g0-expand}
\end{equation}
When this is inserted into the interaction Hamiltonian density in 
Eq.~\eqref{hamiltonian-density}, the term linear in $1/a(t)$ is proportional to the 
contact density operator in Eq.~\eqref{contact-density}.
The term quadratic in $1/a(t)$ is suppressed by $1/\Lambda$
from the additional power of $\bar \lambda_0$. The higher order terms are 
even more highly suppressed.  Thus the interaction Hamiltonian density 
in the zero-range limit can be reduced to
\begin{equation}
{\cal H}_{\rm int}(t) =\
\frac{\bar \lambda_0}{m}\psi_1^{\dagger}\psi_{2}^{\dagger}\psi_2\psi_1
- \frac{1}{4 \pi m} \left( \frac{1}{a(t)} - \frac{1}{\bar a} \right) {\cal C}.
\label{Hint}
\end{equation}
%

%\newpage

%%%%%%%%%%%%%%%%%%%%%%%%%%%%%%%%%%%%%%%%
\section{Matrix Elements of the Contact density operator}
\label{appendix-2}
%%%%%%%%%%%%%%%%%%%%%%%%%%%%%%%%%%%%%%%%

The field theoretic definition of the contact density operator 
in Eq.~\eqref{contact-density} can be expressed as
\begin{equation}
 {\cal C}(\bm r)=\phi^{\dagger}(\bm r)\phi(\bm r),
\label{contact-density-defi} 
\end{equation}
where the {\it contact field} $\phi = \lambda_0 \psi_2 \psi_1$ is a local operator 
that annihilates two atoms at a point.
In the case $a>0$, $\phi(\bm{r})$ has a nonzero amplitude
to annihilate a dimer,
so it can also be referred to as the {\it dimer field}.
The transition matrix element of the contact density operator can be expressed as
%-----------------
\begin{equation}
\langle f | {\cal C}({\bm r})| i \rangle =  
\sum_n\langle f |\phi^{\dagger}(\bm r)|n\rangle\langle n|\phi(\bm r)|i \rangle .
\label{<C>-phi} 
\end{equation}
%----------------- 
A complete set of states $\sum_n | n \rangle \langle n | = 1$
has been inserted between $\phi^\dagger$ and $\phi$.
If only one term in the sum is nonzero, 
the matrix element factors into a matrix element of $\phi$
that involves the initial state 
and a matrix element of $\phi^\dagger$
that involves the final state.

In order to calculate magneto-transition rates in a thermal gas of atoms or dimers,  
one needs to calculate transition matrix elements 
of the contact density operator ${\cal C}(\bm r)$
between two-atom states, which are either a pair of unbound atoms or a dimer.
We will calculate these matrix elements in the Zero-range Model 
defined by the interaction Hamiltonian density in Eq.~\eqref{hamiltonian-density}.
The Feynman rules for the atom propagator  and the 
2-atom--to--2-atom vertex are specified in the
appendix of Ref.~\cite{Braaten:2008uhg}. 
The 2-atom--to--molecule coupling constant $g_0$ should be set to 0.
Using these Feynman rules, the calculation of transition 
matrix elements of the contact density operator can be reduced 
to evaluating Feynman diagrams.

The transition amplitude $A(E_{cm})$ is the amplitude for the transition between 
a pair of atoms in the asymptotic past and a pair of atoms in the 
asymptotic future. It is a function only of the energy $E_{\rm cm}$
of the pair of atoms in their center-of-mass  frame:
\begin{equation}
E_{\rm cm}=E - K^2/4m,
\label{relative-energy}
\end{equation}
where $E$ is their total energy
and $\bm{K}$ is their total momentum. 
(We set $\hbar =1$ in this Appendix.)
The transition amplitude can be calculated by solving the Lippmann-Schwinger equation
shown in Figure~\ref{figure-amp}:
\begin{equation}
 iA(E_{\rm cm})=-i(\lambda_0/m)+(\lambda_0/m)I(E_{\rm cm})A(E_{\rm cm}).
 \label{scattering-amp-2}
\end{equation}
The loop integral $I(E_{\rm cm})$ in the last diagram  in Figure~\ref{figure-amp} is
\begin{equation}
I(E_{\rm cm})=\frac{\lambda_0^2}{m}\int_{\bm q}\frac{1}{q^2-mE_{\rm cm}-i\epsilon},
\label{I-integral}
\end{equation}
where ${\bm q}$ is the loop momentum. Using the expression 
for the bare coupling constant in Eq.~\eqref{coupling},
the solution can be expressed as
\begin{equation}
A(E_{\rm cm})=\frac{4\pi/m}{-1/a+\sqrt{-mE_{\rm cm}-i\epsilon}}.
 \label{scattering-amp-1}
\end{equation}
This amplitude has a pole in the energy at
$E = K^2/4m-1/ma^2$.  The residue of the pole is $-Z_{\rm D}$,
where 
\begin{equation}
Z_{\rm D}=\ 8\pi/m^2a.
 \label{ZD}
\end{equation}

%%%%%%%%%%%%%%%%%%%%%%%%%%%%%%%%%%
\begin{figure}[tbh]
 \centering
 \includegraphics[scale=0.5]{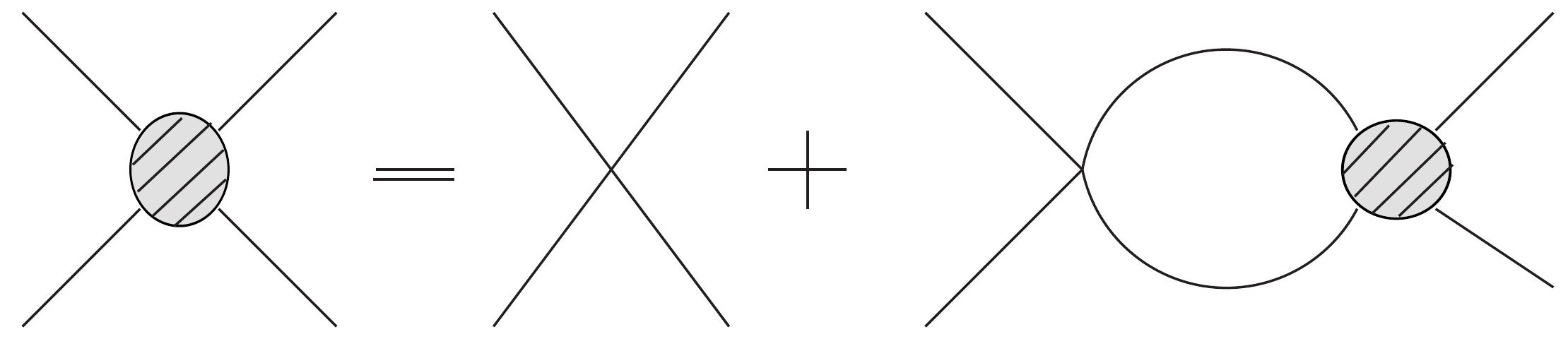}
%\includegraphics[scale=0.5]{Figures/scattering.pdf}
 % scattering.pdf: 0x0 pixel, -2147483648dpi, 0.00x0.00 cm, bb=
 \caption{ 
 The Lippmann-Schwinger equation for the transition amplitude $A(E_{\rm cm})$. 
 The blob represents $iA(E_{\rm cm})$. The vertex factor is $-i\lambda_0/m$.}
 \label{figure-amp}
\end{figure}
%%%%%%%%%%%%%%%%%%%%%%%%%%%%%%%%%%

The standard Feynman rules can be used to calculate matrix elements of 
local operators between states in the asymptotic past and
states in the asymptotic future.  However we need matrix elements
between initial and final states at the same time.
We will use the Feynman rules to calculate matrix elements 
of the contact field operator $\phi^\dagger$
between the vacuum and two-atom states in the asymptotic future.
We will then calculate matrix elements of the 
contact density operator $\phi^\dagger \phi$ between two-atom states 
in the asymptotic future by expressing them in terms of matrix elements 
of $\phi^\dagger$ and $\phi$.

%%%%%%%%%%%%%%%%%%%%%%%%%%%%%%%%%%%%%%%%%%%%%%%
\subsection{Vacuum-to-pair matrix element of $\bm{\phi^{\dagger}}$ }
\label{appendix-vacu-atom}
%%%%%%%%%%%%%%%%%%%%%%%%%%%%%%%%%%%%%%%%%%%%%%%

%%%%%%%%%%%%%%%%%%%%%%%%%%%%%%%%%%
\begin{figure}[tbh]
 \centering
  \includegraphics[scale=0.7]{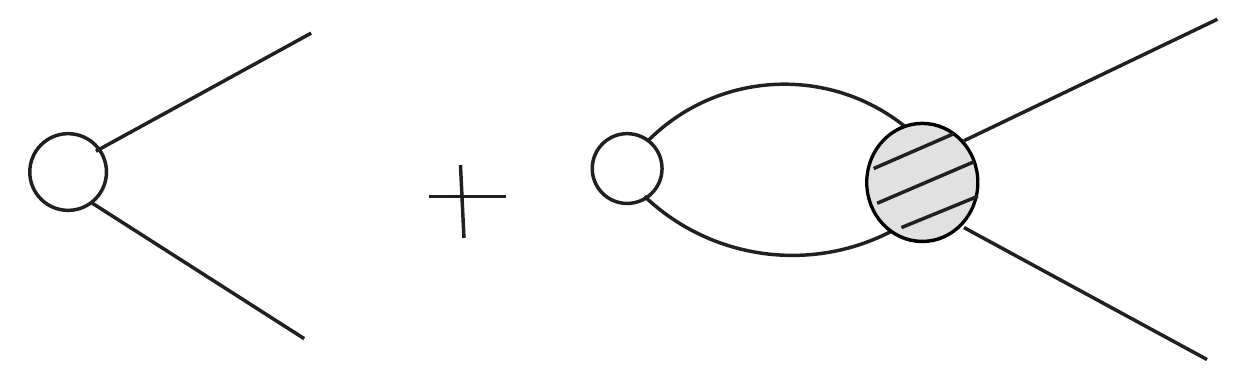}
%\includegraphics[scale=0.7]{Figures/atom.pdf}
 % atom.pdf: 0x0 pixel, 0dpi, nanxnan cm, bb=
 \caption{Feynman diagrams for the matrix element of $\phi^{\dagger}(\textbf{r})$
 between the vacuum and the atom-pair state $|\bm{K}, \bm{k}\rangle$. 
 The open dot represents the $\phi^{\dagger}$ operator, whose Feynman rule is $\lambda_0$. 
 The residue of the pole in the energy in the second diagram 
can be used to determine the matrix element of $\phi^{\dagger}(\textbf{r})$
 between the vacuum and the dimer state $| \bm{k_D}\rangle$.}
 \label{figure-atom}
\end{figure}
%%%%%%%%%%%%%%%%%%%%%%%%%%%%%%%%%%

The matrix element of $\phi^{\dagger}(\bm r)$ between the vacuum $|0\rangle$ 
and an atom-pair state
$|\bm{K}, \bm{k}\rangle$ with total momentum $\bm {K}$ and relative momentum $\bm{k}$
can be calculated from the Feynman diagrams in Fig.~\ref{figure-atom}.
The atoms are on their energy shells with total energy
$E = K^2/4m + k^2/m$.  The matrix element is
\begin{eqnarray}
 \langle {\bm{K}}, {\bm{k}}|\phi^{\dagger}(\textbf{r})|0\rangle&=&
 \lambda_0 \big[1+iI(E_{\rm cm})A(E_{\rm cm}) \big],
 \nonumber\\
 &=&-mA(E_{\rm cm}).
 \label{matrix-element-1-2}
\end{eqnarray}
The energy $E_{\rm cm}$ of the atom pair in their center-of-mass frame
depends only on their relative momentum:
\begin{equation}
E_{\rm cm}=k^2/m.
\label{Ecm-k}
\end{equation}

%%%%%%%%%%%%%%%%%%%%%%%%%%%%%%%%%%%%%%%%%%%%%%%
\subsection{Vacuum-to-dimer matrix element of $\bm{\phi^{\dagger}}$}
\label{appendix-vacu-dimer}
%%%%%%%%%%%%%%%%%%%%%%%%%%%%%%%%%%%%%%%%%%%%%%%

The matrix element of $\phi^{\dagger}(\bm r)$ between the vacuum $|0\rangle$ 
and a dimer state $|{\bm k}_{\rm{D}}\rangle$ with  momentum ${\bm {k}}_{\rm D}$
can be calculated from second Feynman diagram in Fig.~\ref{figure-atom}.
That diagram has a pole in the total energy $E$ of the final-state atoms,
which must be off their energy shells.
At the pole, the center-of-mass energy is equal to the binding energy of the dimer:
$E_{\rm cm} = -1/ma^2$.
The residue of the pole is the product of the desired matrix element and 
$Z_{\rm D}^{1/2}$, where $Z_{\rm D}$ is the residue factor given in
Eq.~\eqref{ZD}.
The matrix element is therefore
\begin{eqnarray}
 \langle {\bm k}_{\rm{D}}|\phi^{\dagger}(\textbf{r})|0\rangle &=&
i \lambda_0 I(E_{\rm cm}) (-Z_{\rm D}) Z_{\rm D}^{-1/2},
\nonumber\\
&=& \sqrt{8\pi/a}.
 \label{phi-dimer}
\end{eqnarray}

%%%%%%%%%%%%%%%%%%%%%%%%%%%%%%%%%%%%%%%%%%%%%%%
\subsection{Pair-to-pair matrix element of $\bm{\phi^{\dagger}\phi}$} 
%%%%%%%%%%%%%%%%%%%%%%%%%%%%%%%%%%%%%%%%%%%%%%%

The matrix element of $\phi^{\dagger}\phi$ between 
the atom-pair states $|{\bm{K}}, {\bm{k}}\rangle$ and $|{\bm{K}^{'}}, {\bm{k}^{'}}\rangle$
in the asymptotic future
can be calculated by inserting a complete set of states between 
$\phi^{\dagger}$ and $\phi$, as in Eq.~\eqref{<C>-phi} .
Since the operator $\phi$ annihilates the initial-state atoms, 
the only term that contributes is the vacuum state.
The  matrix element factors into the vacuum--to--atom-pair matrix element
and its complex conjugate:
\begin{eqnarray}
\langle {\bm{K}^{'}}, {\bm{k}^{'}}|\phi^{\dagger}(\textbf{r})\phi(\textbf{r})|{\bm{K}}, {\bm{k}}\rangle
&=& \langle {\bm{K}^{'}}, {\bm{k}^{'}}|\phi^{\dagger}(\textbf{r})|0\rangle\langle 0|\phi^{\dagger}(\textbf{r})|{\bm{K}}, {\bm{k}}\rangle,\nonumber\\
&=&m^2A^*(E_{\rm cm}) A(E'_{\rm cm}),
 \label{matrix-element-2-1}
\end{eqnarray}
where $E_{\rm cm}$ is given in
Eq.~\eqref{relative-energy} and $E'_{\rm cm}$ is the same expression with
$k$ replace by $k'$. 

The expression for this matrix element given in Ref.~\cite{Braaten:2008uhg}
is incorrect:
the factor  $A^{*}(E_{\rm cm})$ was not complex conjugated.
It is easy to see that this is incorrect by setting the final state equal to the initial state:
$\bm{K}^{'}= \bm{K}$, $\bm{k}^{'} =  \bm{k}$.  
Since the operator is hermitian, the matrix element must be real.
This condition is satisfied by Eq.~\eqref{matrix-element-2-1}.
The error made in Ref.~\cite{Braaten:2008uhg}
was that the matrix element was calculated not between atom-pair states 
at the same time, but between an atom-pair state in the asymptotic past 
and an atom-pair state in the asymptotic future.

%%%%%%%%%%%%%%%%%%%%%%%%%%%%%%%%%%%%%%%%%%%%%%%
\subsection{Pair-to-dimer matrix element of $\bm{\phi^{\dagger}\phi}$}
%%%%%%%%%%%%%%%%%%%%%%%%%%%%%%%%%%%%%%%%%%%%%%%

The matrix element of $\phi^{\dagger}\phi$  between 
the atom-pair state $|{\bm{K}}, {\bm{k}}\rangle$  in the asymptotic future
and the dimer state $|{\bm k}_{\rm D}\rangle$ in the asymptotic future 
can be calculated by inserting a complete set of states between
$\phi^{\dagger}$ and $\phi$, as in Eq.~\eqref{<C>-phi} .
Since the operator $\phi$ annihilates the initial-state atoms, 
the only term that contributes is the vacuum state.
The  matrix element factors into a vacuum--to--dimer matrix element
and the complex conjugate of a vacuum--to--atom-pair matrix element:
\begin{eqnarray}
\langle {\bm k}_{\rm D}|\phi^{\dagger}(\bm r)\phi(\bm r)|{\bm{K}}, {\bm{k}}\rangle
&=&\langle {\bm k}_{\rm D}|\phi^{\dagger}(\textbf r)|0\rangle\langle 0|\phi(\textbf r)|{\bm{K}}, {\bm{k}}\rangle,\nonumber\\
&=&-m\sqrt{8\pi/a}\, A^{*}(E_{\rm cm}),
\label{matrix-2}
\end{eqnarray}
where $E_{\rm cm}$ is given by Eq.~\eqref{relative-energy}.

%%%%%%%%%%%%%%%%%%%%%%%%%%%%%%%%%%%%%%%%%%%%%%%
\subsection{Dimer-to-dimer matrix element of $\bm{\phi^{\dagger}\phi}$}
%%%%%%%%%%%%%%%%%%%%%%%%%%%%%%%%%%%%%%%%%%%%%%%

The matrix element of $\phi^{\dagger}\phi$ between 
the dimer states $|{\bm k}_{\rm D}\rangle$ and $|{\bm k}'_{\rm D}\rangle$
in the asymptotic future
can be calculated by inserting a complete set of states between
$\phi^{\dagger}$ and $\phi$, as in Eq.~\eqref{<C>-phi}.
Since the operator $\phi$ annihilates the initial-state dimer, 
the only term that contributes is the vacuum state.
The  matrix element factors into a vacuum--to--dimer matrix element
and its complex conjugate:
\begin{eqnarray}
 \langle {\bm k}_{\rm D}^{'}|\phi^{\dagger}(\bm r)\phi(\bm r)|{\bm k}_{\rm D}\rangle&=&
 \langle{\bm k}_{\rm D}^{'}|\phi^{\dagger}(\bm r)|0\rangle\langle 0|\phi(\bm r)|{\bm k}_{\rm D}\rangle,\nonumber\\
 &=&8\pi/a.
 \label{matrix-dimer-3}
\end{eqnarray}

\end{acknowledgments}

%\newpage

\end{document}